\documentclass[preprint2]{mn2e}
\usepackage{graphicx}
\usepackage {amsmath,amssymb}
\usepackage{color}
\newcommand{\hi}{{\rm H}{\textsc i }}
\def\h2{\rm{H_2}}
\def\fHI{f_{\rm{HI}}}
\def\fh2{f_{\rm{H_2}}}
\def\mHI{M_{\rm{HI}}}
\def\mHIcand{M_{\rm{HI,cand}}}
\def\mh2{M_{\rm{H2}}}

\def\Sh2{\Sigma_{\h2}}

\def\ms{M_{\odot}}

\def\HIs{M_{\rm{HI}}/M_*}

\begin{document}
\title{An HI View of  Galaxy Conformity: HI-rich Environment around HI-excess Galaxies}

\author[Jing ~ Wang et al.]{Jing Wang$^{1}$\thanks{Email: j.wang@csiro.au}, Paolo Serra$^{1}$, Gyula I. G. J\'ozsa$^{2,3,4}$, B\"arbel Koribalski$^{1}$,\\ 
\newauthor  Thijs van der Hulst$^{5}$, Peter Kamphuis$^{1}$,  Cheng Li$^{6}$ , Jian Fu$^{6}$, Ting Xiao$^{6}$, \\
\newauthor Roderik Overzier$^{7}$, Mark Wieringa$^{1}$, Enci Wang$^{6}$\\
$^1$Australia Telescope National Facility, CSIRO Astronomy and Space Science, PO box 76, Epping, NSW 1710, Australia\\
$^2$SKA South Africa Radio Astronomy Research Group, 3rd Floor, The Park, Park Road, Pinelands, 7405, South Africa\\
$^3$Rhodes University,  Department of Physics and Electronics, Rhodes Centre for Radio Astronomy Techniques \& Technologies,\\
 PO Box 94, Grahamstown, 6140, South Africa\\
$^4$Argelander-Institut für Astronomie, Auf dem H\"ugel 71, D-53121 Bonn, Germany\\
$^5$University of Groningen,  Kapteyn Astronomical Institute, Landleven 12,  9747 AD, Groningen, The Netherlands\\
$^6$Key Laboratory for Research in Galaxies and Cosmology, Shanghai Astronomical Observatory, Chinese Astronomical Society, \\
 80 Nandan Rd, Shanghai 200030, China\\
$^7$Observat\'orio Nacional, Ministry of Science, Technology, and Innovation, Rio de Janeiro, Brazil \\
  }
\date{Accepted 2014 ???? ??
      Received 2014 ???? ??;
      in original form 2014 January}

\pubyear{2014}
\maketitle

\begin{abstract}
Using data taken as part of the Bluedisk project we study the connection between neutral hydrogen (\hi) in the environment of spiral galaxies and that in the galaxies themselves. We measure the total \hi mass present in the environment in a statistical way by studying the distribution of noise peaks in the \hi data cubes obtained for 40 galaxies observed with WSRT. We find that galaxies whose \hi mass fraction is high relative to standard scaling relations have an excess \hi mass in the surrounding environment as well. 
Gas in the environment consists of gas clumps which are individually below the detection limit of our \hi data. These clumps may be hosted by small satellite galaxies and$\slash$or be the high-density peaks of a more diffuse gas distribution in the inter-galactic medium. We interpret this result as an indication for a picture in which the ${\rm H}{\textsc i}$-rich central galaxies  accrete gas from an extended gas reservoir present in their environment. 

\end{abstract}


\section{Introduction}
Gas accretion is an important process in the evolution of galaxies. 
Galaxies like the Milky Way would run out of their gas in less than 1 Gyr given their current star formation rate unless their gas reservoir is continuously replenished (Kennicutt 1983). The source of this fresh gas is unknown but the majority is unlikely to be contained in gas-rich satellite galaxies because they can contribute at most 10 percent of the required gas mass (Kauffmann et al. 2010; Di Teodoro \& Fraternali 2014, ). Likewise,  high-velocity \hi clouds  contribute 3-5 times less gas than required to sustain the Milky Way star formation (Richter 2012).
 A potentially more substantial source of fresh material is the cooling flow of low-metallicity, ionised gas in a galaxy's halo (Shull et al. 2009). This may be related to the "fountain" gas which is driven to a height of a few kpc from the disc by star formation, and may trigger a gas inflow rate very close to the star formation rate (Fraternali et al. 2013).
 Recent theoretical studies suggest that the fountain driven gas naturally produces an outside-in shrinking galactic disk (Elmegreen et al. 2014), which is consistent with what  
is observed for nearby dwarf galaxies (Zhang et al. 2012). However, it does not explain the observed inside-out growth of more massive disk galaxies (J. Wang et al. 2011), or their chemical evolution (Spitoni et al. 2013), which have been both explained in terms of a cosmological radial gas accretion (Kauffmann 1996, Spitoni et al. 2013). 

In the framework of the standard $\Lambda$CDM cosmology, two modes of (radial) gas accretion occur.  Gas entering the potential well of a galaxy gets shocked to the virial temperature of the dark matter halo at an early epoch of infall, and then gradually cools and falls onto the central galaxy (Rees \& Ostriker 1977; Silk 1977; Binney 1977, White \&Rees 1978). 
 Gas that has never been shock heated  close to virial temperature falls onto the galaxy from outside the virial radius along filamentary structures (e.g. Kere{\v s} et al. 2005). This latter mode is termed ``cold mode'' accretion, and is believed to be the dominant way of gas accretion for galaxies
in low mass dark matter halos (Kere{\v s} et al. 2005, Dekel \& Birnboim et al. 2006). Numerical studies characterise the filaments of cold mode accretion as cool and clumpy clouds surrounded by  ionised gas (Kere{\v s} \& Hernquist 2009).

There have been many observational attempts to trace the cold mode accreting gas around galaxies.  Most of the evidence comes from observing ionised gas at intermediate to high redshift.
QSO absorption lines in Mg II  with an inflowing velocity feature have been found to be prevalent at a distance of tens of kpc around star forming galaxies (e.g. Giavalisco 2011, Kacprzak et al. 2012). At least one of them is observed to have  a filamentary structure (Rauch et al. 2011). Furthermore, metal-poor Lyman limit systems are observed to have characteristics expected for cold accretion (Lehner et al. 2013).  

Tracing the cold-mode accreting gas in the neutral phase in galaxies at low redshift has been difficult. 
\hi clouds that form a filamentary structure have been found around some nearby galaxies (e.g., Shostak \& Skillman 1989; Oosterloo et al. 2007; de Blok et al. 2014, for a review see
Sancisi et al. 2008) and in the Milky Way (HVCs, Putman et al. 2002) but are more likely to trace interaction with companions and star formation feedback. None of them has been confidently demonstrated to be tracing cold-mode gas accretion.
Although cold-mode accretion is believed to produce the highly clumpy galactic disks observed at high redshift (Agertz et al. 2009), we do not find an especially clumpy morphology for the ${\rm H}{\textsc i}$-rich galaxies when compared to control galaxies at low redshift (J. Wang et al. 2013, 2014).
This may be because at low redshift the accreting gas gets slowed down and blends better with the hot halo gas around the galaxies (Putman et al. 2012).
Hence, it may be better to search for the cold-mode accreting neutral gas well outside the galactic disks, by searching for signal from a large volume around the central galaxies. 
Here we attempt to perform exactly this search by analysing the data taken as part of the Bluedisk project (J. Wang et al. 2012) in a special way.
  
 Previous analysis of the same Bluedisk dataset already shows that the gas richness of a galaxy is linked to the gas richness of its environment. In particular, the satellites of abnormally ${\rm H}{\textsc i}$-rich spirals are themselves abnormally \hi rich (E. Wang et al. 2015). This result may be seen as a new manifestation of the ``conformity phenomenon'', whereby the colours of satellites correlate with the colour of central galaxies (Weinmann et al. 2006, Kauffmann et al. 2010, W. Wang et al. 2012). The work of E. Wang et al. (2015) was based on individual \hi detections in the Bluedisk data cubes and, therefore, is limited to galaxies with $\mHI>10^8~\ms$. As previously pointed out by Kauffmann et al. (2010), because the galactic conformity usually extends to scales far beyond dark matter halos,  it implies that the satellite galaxies are probably tracing the underlying smooth gas reservoir that is available for fuelling the central galaxies.
 If this hypothesis is true,  the conformity behaviour should still hold when it is examined at low luminosity levels (i.e. by tracing the less bright satellites or clouds in the general environment that are potentially far more numerous than the brighter satellites). 

 Motivated by these results we perform a new analysis of the Bluedisk data. Our aim is to trace the faint gas distribution that may surround galaxies below the detection limit of the \hi data. Our method consists of adding up the flux of (both positive and negative) noise peaks in the \hi data cubes out to a radius of 16 arc min (typically 500 kpc), trying to detect a statistical \hi signal (i.e. an excess of positive detections) and studying whether its properties correlate with the \hi content of the central galaxy.

 The paper is organised as follows. In section 2, we describe the Bluedisk data used in this analysis.  In section 3, we introduce our new method  to measure a total  \hi mass in faint systems around primary galaxies through examining the noise peaks. 
In section 4, we discuss the origin of the \hi mass measured from an analysis of the noise peaks. We find that the excess \hi mass fraction in the primary galaxies is correlated with an excess \hi mass in the surrounding environment.  In section 5, we discuss the implication of our results for cold mode gas accretion in galaxies. A $\Lambda$CDM cosmology with $\Omega_{m}=0.3$, $\Omega_{\lambda}=0.7$ and $h=0.7$ is assumed throughout the paper.

\section{The sample}
\subsection{The Bluedisk dataset}
The Bluedisk project aims at searching for clues about gas accretion and galactic disk formation in the local universe. Details about the sample, data processing and analysis can be found in Paper I and II, and here we only review the most relevant information.

 The sample consists of 50 massive galaxies (M$_*>10^{10} \ms$) with a redshift between 0.018-0.03 (corresponding to a luminosity distance of 70-130 Mpc).  From the atlas in Paper I, most of them are blue spiral galaxies.
The target fields are  covered by the Sloan Digital Sky Survey (SDSS, Abazajian et al. 2009), the Galaxy Evolution Explorer (GALEX) imaging survey (Martin et al. 2005), and the Wide-Field Infrared Survey Explorer  (WISE, Wright et al. 2010), allowing a multi-wavelength analysis. 
Radio 21 cm synthesis data were obtained with the Westerbork Synthesis Radio Telescope (WSRT), with the main goal of observing the \hi emission line. 
In this paper, we make use of the \hi  data cubes optimised for high sensitivity at lower resolution, which are produced with Robust 0 weighting and 30$"$ tapering.   The data cubes have a typical rms of 0.37 mJy beam$^{-1}$, a typical beam size of 38$'' \times 36''$ and a velocity resolution of 12.3 km/s (after Hanning smoothing), which provides sufficient sensitivity to detect point sources  down to an \hi mass of a few $10^8\ms$.  The primary beam has a FWHM (full-width-at-half-maximum) of $\sim0.5\degr$, corresponding to a $\sim1$ Mpc sky region around the target galaxies,  hence the data is very suitable for studying the relatively large-scale \hi environment around the central sources. The details of the primary beam correction are discussed in E.Wang et al.(2015).

Following Paper I, we use the 42 isolated target galaxies for this study. Hence, the local environment is different from groups or clusters, where the dominant process is gas stripping rather than accretion. We further exclude two galaxies  (galaxy 14 and 27) with the lowest redshift, so that we have a narrow redshift range of 0.023--0.03 (a luminosity distance of 100-130 Mpc).
We call these 40 target galaxies  ``primary galaxies''.  

\begin{figure}
\includegraphics[width=8cm]{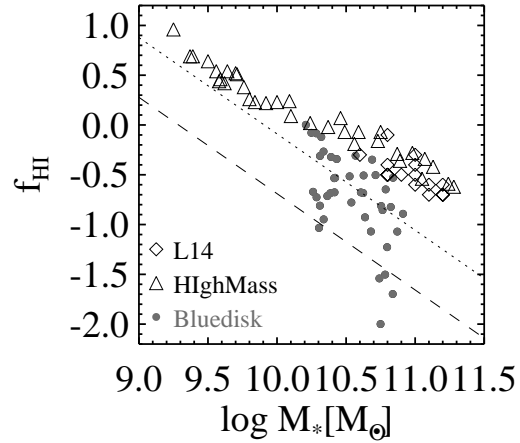}
\caption{ \hi mass fraction of the Bluedisk sample (dark grey dots), the HIghMass sample (triangles), and the sample from Lemonias et al. (2014) (L14, diamonds). The dashed line shows the median relation between $\fHI$ and stellar mass for general galaxies, taken from Catinella et al. (2010). The dotted line is the dashed line 0.6 dex upward, showing the original division between the Bluedisk high $\fHI$ and control galaxies.}
 \label{fig:sample} 
\end{figure}

These galaxies have a broad distribution in \hi mass fraction $\fHI=\log \HIs$, ranging from lower than -2 to higher than 1 (Figure~\ref{fig:sample}). Compared to other samples featuring ${\rm H}{\textsc i}$-rich galaxies (like the samples from Lemonias et al. 2014  and the HIghMass project, High \hi Mass, Huang et al. 2014), the Bluedisk sample has the advantage of being able to compare the most \hi rich galaxies with those of normal \hi content. As we will show later in Section~\ref{sec:conformity}, this is crucial for producing our main results.

As mentioned before, the aim of the paper is to investigate the connection between the \hi in the (central) primary galaxies and the \hi in the environment. In addition to $\fHI$, we also use a few other parameters to quantify the \hi richness of primary galaxies. 
Using the GALEX Arecibo SDSS Survey (GASS, Catinella et al. 2010),  Catinella et al. (2013, {\bf C13 for short hereafter}) calibrated a photometric estimator of $\fHI$ in galaxies with NUV$-r<4.5$ (which is also the colour range of the Bluedisk sample). The estimator is a combination of the NUV$-r$ colour and the mass surface density,  reflecting the connection between star formation rate and cold gas content. The difference between the observed  and the estimated $\fHI$, which we denote as $\Delta_{C13}\fHI$, measures the excess gas, and can be used to indicate \hi richness. 
Xiao et al. (in prep)  further add the optical concentration parameter and stellar mass to the estimator, giving more control on the internal structure of galaxies. We also use the difference from this  4-parameter (4p) estimator ( $\Delta_{4p}\fHI$) as a measure of  \hi richness.

\section{Method}
 As discussed in the introduction section, the conformity behaviour of central and bright satellite galaxies can be interpreted in terms of an underlying extended reservoir of cold gas. If this hypothesis is correct,  we expect to detect conformity also at lower \hi mass levels than probed by E. Wang et al. (2015), in a regime dominated by very faint satellite galaxies and$\slash$or gas clouds in the inter-galactic medium.
In this section we introduce our technique to  detect this
  faint \hi gas,  in the environment of the Bluedisk galaxies.
      
The method entails an analysis of the noise peaks (both positive and negative) in {\bf the cleaned data cubes} in order to obtain a statistical detection. Because the noise is distributed symmetrically around 0 (both the median and mean background of a cube are typically within $\pm$10$^{-5}$ mJy beam$^{-1}$, while the rms of a cube is $\sim$0.4 mJy beam$^{-1}$), a positive noise peak will statistically be cancelled out by a negative noise peak of the same level. 
If in addition to noise there is a collection of very faint \hi objects, then these will add a slight positive signal to the noise, that will elevate the positive noise peaks or diminish
negatiive noise peaks. This bias is detectable by a careful analysis of the distribution of the noise values throughout the cube. We do this by selecting all (positive and
negative) noise peaks above 4 $\sigma$ in the area of interest, after removing the signal that has been reliably detected in the source finding procedure ( see section~\ref{sec:scfinding}).

Selecting noise peaks above a certain threshold has an important advantage compared to integrating over all voxels within a large area.  Bright HI emission as well as diffuse faint HI emission in the data cubes is cleaned by means of clean masks (Paper I). The clean masks do not include the faint noise peaks used for our statistical analysis here, and these are therefore not cleaned. 
 Because the integral over the {\bf dirty beam} over a large area goes to zero it becomes very difficult to determine fluxes over extended areas without proper deconvolution. This is inherent to the sidelobe structure of the PSF. Any blind measurement of the total flux over a large area is therefore fraught with problems.
 However, by pre-selecting noise peaks, we analyse only the central part of the PSF and exclude any sidelobes. We are therefore unable to detect the \hi mass of a very extended, smoothly distributed medium, (this is a consequence of using an interferometer with limited short baselines), but we are able to detect small scale faint emission using the method outlined here.
 
In the following sections, we describe our procedure in details. We improve the  data quality by flattening the spectral baselines (section~\ref{sec:flattening}).
We use the source-finding application SoFiA (Serra et al. 2015) to extract all signal
exceeding 4 $\sigma$. After removing all bright, reliable detections (section~\ref{sec:scfinding}), we are left with
``noise'' peaks  which we analyse in order to detect a statistical signal.  
We perform a series of careful tests to examine the robustness of the accumulative signal against various data reduction artefacts (section~\ref{sec:testing}).

\subsection{Flattening the spectral baselines}
\label{sec:flattening}
In order to improve the quality of the \hi data cubes we fit second-order polynomial functions in moving windows along each line of sight. This allows us to obtain global
    functions describing the shape of the spectral baselines without specifying their
    functional form. In practice, we convolve each spectrum with the
    Savitzky-Golay filter (S-G, Savitzky \& Golay 1964), an analytical solution for the
    local fitting process.  We adopt a S-G filter with a half width of 70 channels (840 km/s),  
    wider than the velocity width of all the  galaxies in our data cubes. This ensures that the smoothed
    spectrum describes the large-scale shape of the baseline only. 
    In Figure 2 we show the shape of the filter and demonstrate the convolution
    effect on simple functions, including the presence of edge effects. Within our
    analysis range of $\pm$500 km/s around the spectrum centre (channels 34 to 114),
    the S-G filter conserves the shape of the functions with a maximum deviation
    of less than 15\%.

Before convolving the \hi spectra with the S-G filter we mask channels that belong to one of the \hi detections presented in Paper I (see also Sec. 3.2 below).
 We refer to these channels as the initial flagging set, and replace them with values interpolated from a first order polynomial fitting to the un-flagged part of spectrum. 
We calculate $\sigma_m$, the median absolute deviation  of the original spectrum from the convolved spectrum,
 and update the mask by including all channels that deviate by more than $\pm$3 $\sigma_m$ from the convolved spectrum.
  A flagged channel is replaced by the value interpolated from a first order polynomial fit to the un-flagged part of the whole spectrum, if it is in the initial flagging set, or in a flagged region that is wider than 10 channels and  less than 40 channels from one end of the spectrum (potential bright sources that lie near the edge of the spectrum and cause incorrect continuum over-subtraction during the former data reduction).  Otherwise the flagged channel (which is not likely to belong to a bright source) gets replaced by a first order polynomial interpolation from  nearby channels. 
We repeat the above steps of flagging and convolution until $\sigma_m$ (with typical value of a few times 0.1 mJy beam$^{-1}$) varies by less than 10$^{-5}$ mJy beam$^{-1}$, or after 10 iterations. 
Figure~\ref{fig:subcont} shows that this procedure models a variety of spectral baseline shapes successfully.

The final step of this procedure consists of combining all spectral baseline models into individual cubes. We smooth these cubes with a gaussian kernel with a FWHM of 2 pixels  in the Ra-Dec direction in order to reduce the fluctuation between adjacent pixels (this does not significantly modify the PSF (which has a typical FWHM of 3.7 pixels).  Finally, we subtract the resulting spectral baseline model cube from the original cube.
After this subtraction, the median rms of each data cube drops very slightly from 0.367 to 0.363 mJy beam$^{-1}$ (and from 0.372 to 0.362 mJy beam$^{-1}$ for the inner channel range 34 to 114). Hence while flattening the baselines, the method does not significantly change the global noise properties of the cube,  
In Figure~\ref{fig:projection}, we compare two data cubes, using the same position-velocity cut,  one as obtained before the additional continuum subtraction, and one as obtained after the additional continuum subtraction. While the noise features look similar, the continuum subtraction is obviously improved.

\begin{figure}
\includegraphics[width=8cm]{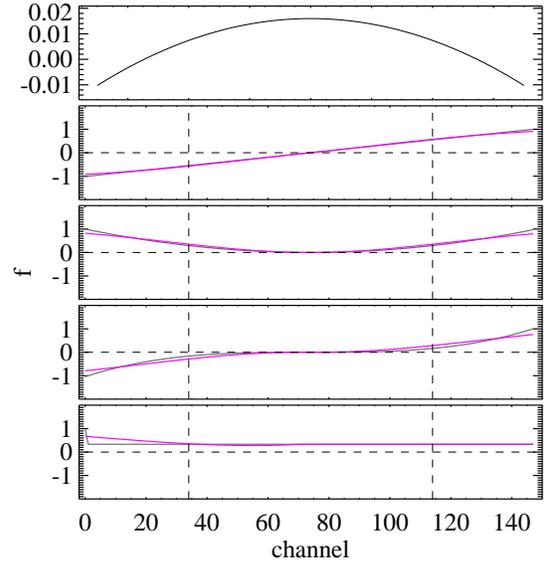}
\caption{The S-G filter (top row)  and  examples of the convolution effect on simple functions.  The area under the S-G filter in the top row is normalised.
From the second to the bottom row, the dark grey curves are linear, second order polynomial and third order polynomial functions and a flat line with a spike on one side. The magenta curves are the result of the grey curves convolved with the S-G filter. The horizontal dashed lines mark the y axis position of 0, and the vertical lines mark the channels  34 and 114.}
 \label{fig:sgfilter}
\end{figure}

\begin{figure}
\includegraphics[width=8cm]{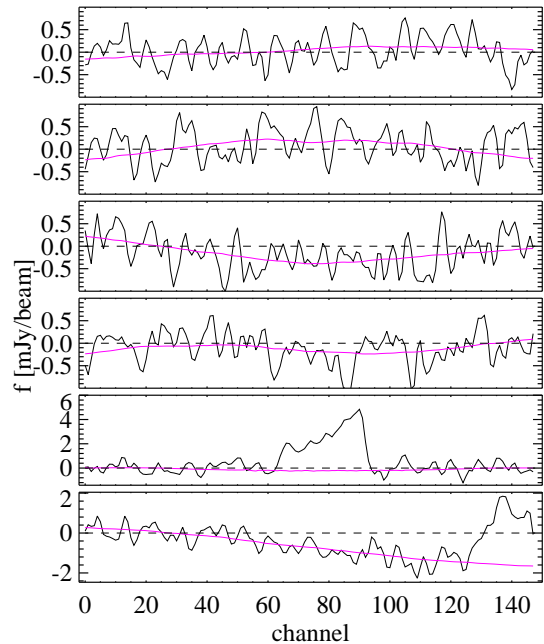}
\caption{Examples of spectra before (black lines) and after applying the S-G filter (magenta lines). The dashed lines mark the zero level.}
 \label{fig:subcont}
\end{figure}

\begin{figure}
\includegraphics[width=7.5cm]{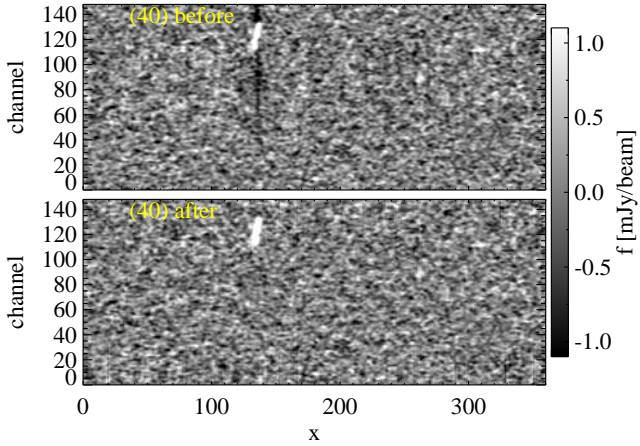}
\caption{ An example of an x-z projection from an original \hi cube (top) and that after applying the improved continuum subtraction discussed in section~\ref{sec:flattening} (bottom). The displayed grey scale ranges from -3 to 3 $\sigma$ around the mean of the cube.  }
 \label{fig:projection}
\end{figure}

\subsection{Source finding and statistical detection of HI in the environment}
\label{sec:scfinding}
We extract sources (both real detections as well as noise peaks to be used for our statistical analysis) from the improved \hi data cubes \footnote[1]{The data cubes are not corrected for primary beam attenuation. The \hi mass of sources are corrected for the primary beam attenuation after they are extracted. } using the source detection software SoFiA\footnote[2]{https://github.com/SoFiA-Admin/SoFiA}. 
We refer  to Serra et al. (2015) for a detailed description of the software.  Here, we only outline the most relevant steps  and describe the key parameter choices. 

SoFiA first convolves the data with various smoothing kernels and selects voxels with absolute values above a certain threshold.  We use smoothing kernels with a width of 0  and 3 pixels in the projected sky direction (roughly the size of the synthesised beam), and a width of 0, 3 and 5 pixels (corresponding to 0, 36 and 60 km$/$s) in the velocity direction (smoothing  with a kernel with width of 0 is equivalent to no smoothing). Each channel of the cube is weighted according to the  noise, so that noise variations along the velocity axis are removed. This step is very useful to suppress the detection of weak RFI (radio frequency interference) residuals. 
We adopt a detection threshold equal to 4 $\sigma$. We note that the clipping is performed on the absolute values of the voxels in the filtered data cubes, meaning that our initial source list contains sources with negative total flux. 
These parameter settings for SoFiA are chosen after several tests and provide a good balance between suppressing the noise and gathering sufficient statistics for the following analysis..

SoFiA calculates a reliability index $R$ for each source with positive total flux (Serra et al.2012, 2015). We chose 0.01 Jy beam$^{-1}$ as a minimum total flux and R$>$0.99 to classify a positive detection as a reliable source. We will refer to the remaining positive and negative detections (the noise peaks) as the ``candidates'' hereafter.  While the division between reliable sources and candidates is not strict, we will see in section~\ref{sec:matchopt} that the candidates close to the threshold of reliable sources do not dominate the total signal in candidates.  

Using these selection criteria, besides the 40 primary galaxies, SoFiA detects in total 59,757 sources from the 40 cubes, of which, using the scheme as described, 137 are classified as reliable. 
42 of the reliable sources lie within a projected distance ($D_{proj}$) of 500 kpc and a systematic velocity  difference ($\Delta V_{sys}$) of 500 km$/$s around the primary galaxies.  95$\%$ of them have $\mHI>1.62\times10^8\ms$.  All of them can be associated with an optical galaxy down to a $r$ band magnitude of 22 by requiring the centre to be offset by less than the size of the \hi beam. The connection between primary galaxies and these reliable sources (satellite galaxies) is studied in detail in E. Wang et al. (2015). One of the key findings is that the satellites of ${\rm H}{\textsc i}$-excess galaxies ($\Delta_{C13}\fHI>0$; see Sec. 2) have higher \hi excess than the satellites of ${\rm H}{\textsc i}$-normal galaxies ($\Delta_{C13}\fHI<0$). Here we aim to test whether such a connection between the \hi content of galaxies and that of their environment holds also when studying the fainter, candidate \hi sources. 

We find 6,369 candidates that lie within a $D_{proj}$ of 100 to 500 kpc and $V_{sys}$ distance of 500 km$/$s from the primary galaxies, and these will serve as the main sample for analysis in what follows. 
We exclude the inner 100 kpc region because some of our \hi rich galaxies have large \hi disks extending to $\sim$ 100 kpc. The radial range is always confined within the full-width-half-power of the primary beam. Most of the candidates have \hi masses (absolute values for the negative candidates) within the range of  $4.0\times10^6\ms$ and $1.4\times10^8\ms$ (5 and 95 percentiles of the distribution).

The ensemble of all positive and negative candidate detections results in a net positive \hi signal in the environment around Bluedisk galaxies. We show this by calculating for each cube the accumulative \hi mass ($\mHIcand$) by summing up all negative and positive candidates in the volume defined above. 
To further improve the signal-to-noise ratio, we calculate the average $\mHIcand$ of the 40 cubes. We calculate the error bars as a combination of the  variation over the data cubes (var1) and the mean variation within a data cube (var2). The variation over the data cubes (var1) is calculated through a standard bootstrapping procedure. Taking the whole sample of 40 cubes for example, we randomly select 40 cubes from the sample (repetition allowed) and calculate the average $\mHIcand$. We repeat this step for 1000 times and obtain a distribution consisting of 1000 values of average $\mHIcand$. var1 is calculated as the variation of this distribution.  The final error is calculated as  $\sqrt{var1+\overline{var2}}$.
We derive an average $\mHIcand$ of $\sim1.1\times10^8\ms$ for the 40 cubes within the whole analysis volume. 

 The question is, however, whether and how significantly $\mHIcand$ is  affected by remaining artefacts in the data. Asymmetrically distributed noise  can come from weak RFI  residuals that are not completely removed during data reduction, and can also be produced in the continuum subtraction and cleaning processes.  
Hence in the following  sub-sections we undertake a series of tests to identify and quantify these effects,  before moving on to the scientific interpretation in section~\ref{sec:result}.

\subsection{Influence of synthesis data reduction on the mass in candidates}
\label{sec:testing}
\begin{figure*}
\includegraphics[width=14.cm]{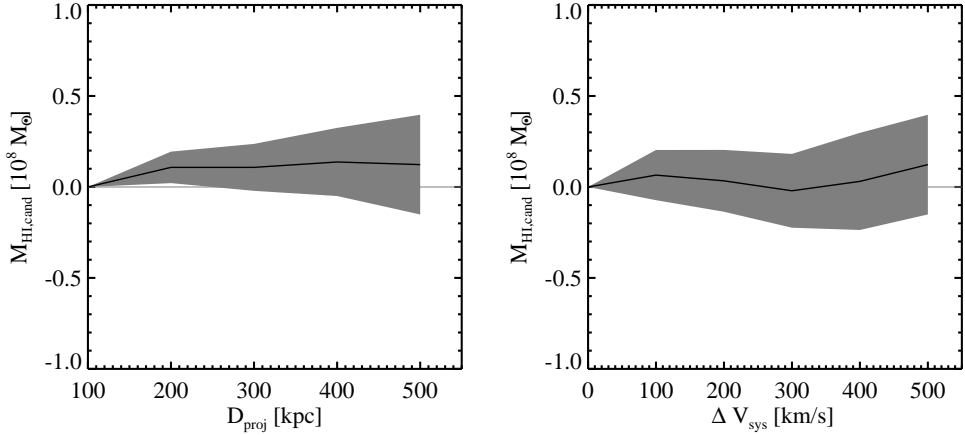}
\caption{The average accumulative $\mHIcand$  from stokes Q cubes as a function of projected distance (left) and systematic velocity distance (right) to the primary galaxies.   The solid black lines show the average profiles around each primary galaxy, and the grey shades show the errors.  }
 \label{fig:mhidistr_qq}
\end{figure*}

We begin this series of tests by verifying that the white noise in the data does not accumulate to a significant net signal. We test this by applying the same source finding and  $\mHIcand$ measurements presented in section~\ref{sec:scfinding} to the Stokes Q  cubes. Since no polarised signal in the line data is expected, the result is expected to be 0 unless indeed a positive accumulative  noise signal is present. Figure~\ref{fig:mhidistr_qq} confirms that $\mHIcand$ of the Stokes Q cubes is indeed close to 0. 

In addition, we visually inspect the data cubes carefully in various projections. The well trained eye is able to spot and identify systematics in the data. Because the WSRT is a linear E-W array, bright RFI residuals  show patterns of parallel stripes and bright cleaning residuals show patterns of elliptical rings around bright sources on the Ra-Dec projections. Improper continuum subtraction results in negative regions on  the Ra-velocity or Dec-velocity projections (see Figure~\ref{fig:projection}). 
We find no obvious cleaning or continuum subtraction residuals.  We find some very weak underlying RFI patterns in a few channels, but none of our \hi candidates are associated with any of those. This is partly because the source finder takes into account noise variations within cubes (including those caused by residual RFI; section~\ref{sec:scfinding})

We continue our search for possible sources of systematic error in our measurement of $\mHIcand$ by investigating the effect of incorrect cleaning and continuum subtraction. Any systematics related to clean residuals {\bf (i.e. the presence of residual side lobe flux in the HI cubes)} should correlate with a higher cleaned flux or a larger cleaned region. This opens the possibility for the following tests.

\begin{figure*}
\includegraphics[width=16.cm]{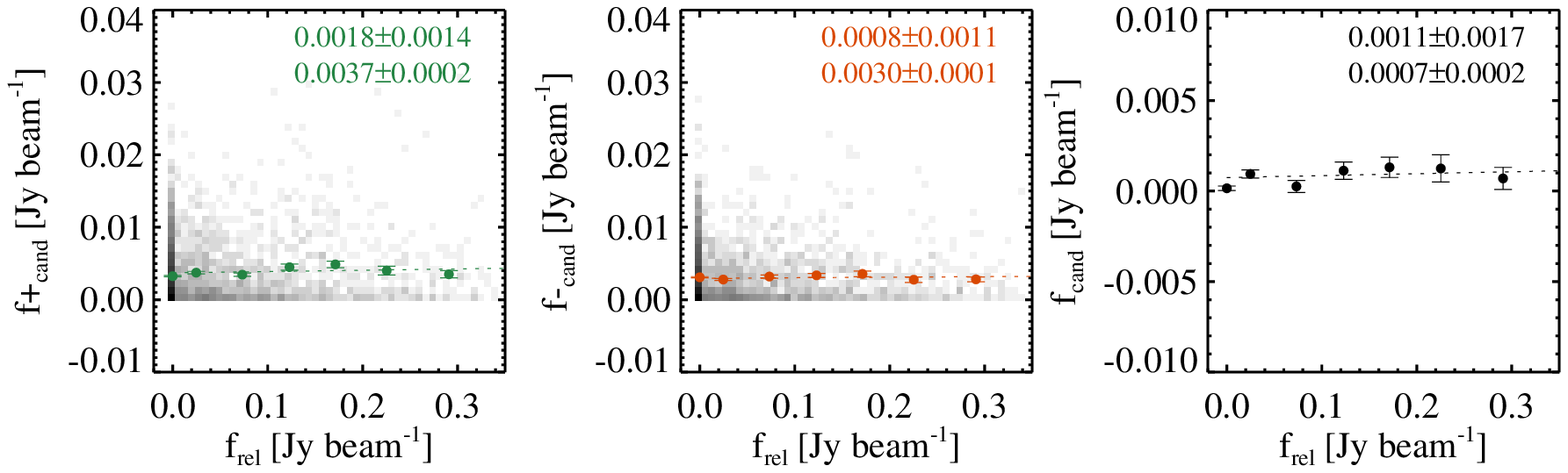}
\includegraphics[width=16.cm]{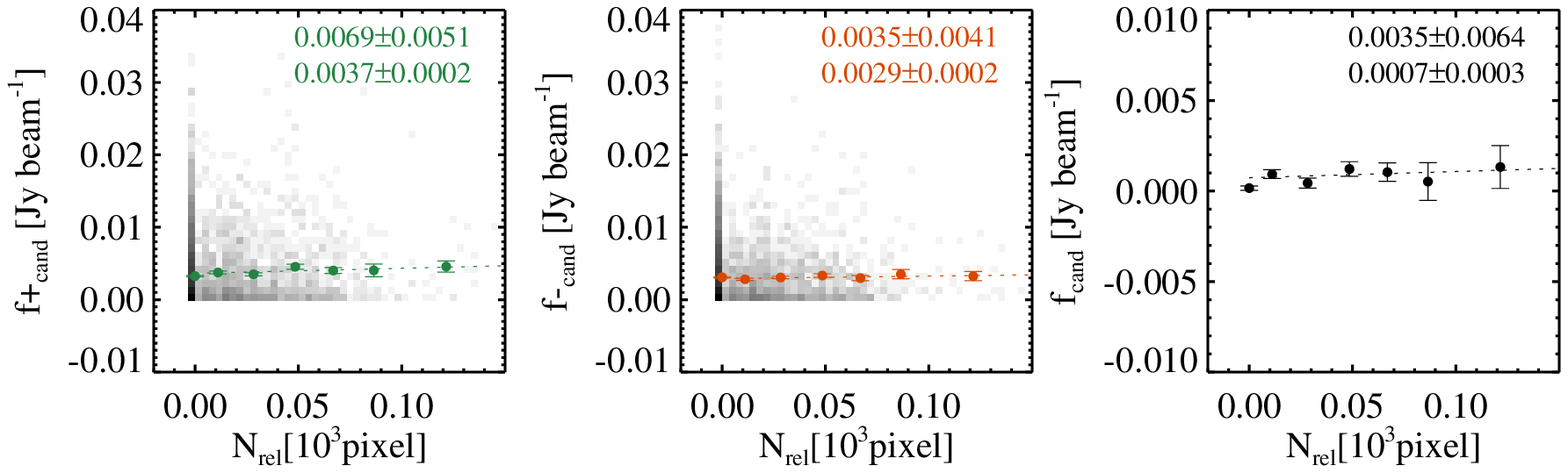}
\caption{  Relations of fluxes in candidates and the reliable sources in channels. $f+_{cand}$ is the total flux of positive voxels in candidates, and $f-_{cand}$ is the absolute total flux of negative voxels in candidates. $f_{rel}$ and N$_{rel}$ are the total flux and voxel number in reliable sources. In each  panel, the grey scale map shows the number density of data points,  the dots with error bars represent the mean values in each x-axis bin,  and the dotted line shows a linear fit to the dots, with slope and intercept of the fit denoted (in the first and second row, respectively) on top of the panel.}
 \label{fig:error_cleaning} 
\end{figure*}

  We calculate the sum of all positive voxels belonging to candidates and the absolute sum of all the negative voxels belonging to candidates, and we refer to them as $f+_{cand}$ and $f-_{cand}$.  We calculate the total flux in the reliable sources and the total flux in the candidates, and refer to these as $f_{rel}$ and $f_{cand}$ respectively.  We calculate $f+_{cand}$, $f-_{cand}$, $f_{cand}$ and $f_{rel}$ for each channel map within 500 km$/$s from the systemic velocity and 500 kpc from the position of the primary galaxy.  These are the velocity range and sky region within which we perform our analysis.
  The top row of Figure~\ref{fig:error_cleaning} shows that there is no correlation between $f-_{cand}$ and $f_{rel}$, suggesting that clean residuals do not significantly affect the noise (negative voxels) of the data cube. There is a very weak relation between $f+_{cand}$ and $f_{rel}$, resulting in a systematic linear relation of $f_{cand}$ increasing slightly as a function of $f_{rel}$. We argue that part of the relation reflects galaxy-environment connections rather than artefacts, as we will show later.  If we attribute the relation fully to clean artefacts around reliable sources,  on average an \hi mass of 0.11$\times10^8\ms$ related to clean artefacts is present in each data cube, which is an order of magnitude lower than the average \hi mass contained in the candidates ($\mHIcand\sim1.1\times10^8~\ms$, section~\ref{sec:scfinding}).   
 In the bottom row of Figure~\ref{fig:error_cleaning}, we perform a similar analysis as in the top row but replace $f_{rel}$ with $N_{rel}$, the number of voxels in the reliable sources. We get similar trends, and the linear relation between $f_{cand}$ and $N_{rel}$ suggests that on average a maximum positive bias in \hi mass of 0.07$\times10^8\ms$  is associated with the cleaning of reliable sources in a data cube. 
 These tests demonstrate that the cleaning process may indeed leave a small amount of positive residual flux, but it does not significantly affect  $\mHIcand$ of the candidates.
 
Are there faint systematic residuals from continuum subtraction that can not be directly caught by eye? We extract continuum sources from the continuum maps, with a 3 $\sigma$ detection threshold. These sources have a 5 percentile of 0.94 mJy in the flux distribution.
We find that 99$\%$ of the flux in candidates comes from sight lines without detectable continuum sources. The fraction is higher than 95\% for both positive and negative candidates, if we count them separately. 
Hence continuum subtraction residuals are not likely to significantly affect  $\mHIcand$.

\begin{figure*}
\includegraphics[width=4.cm]{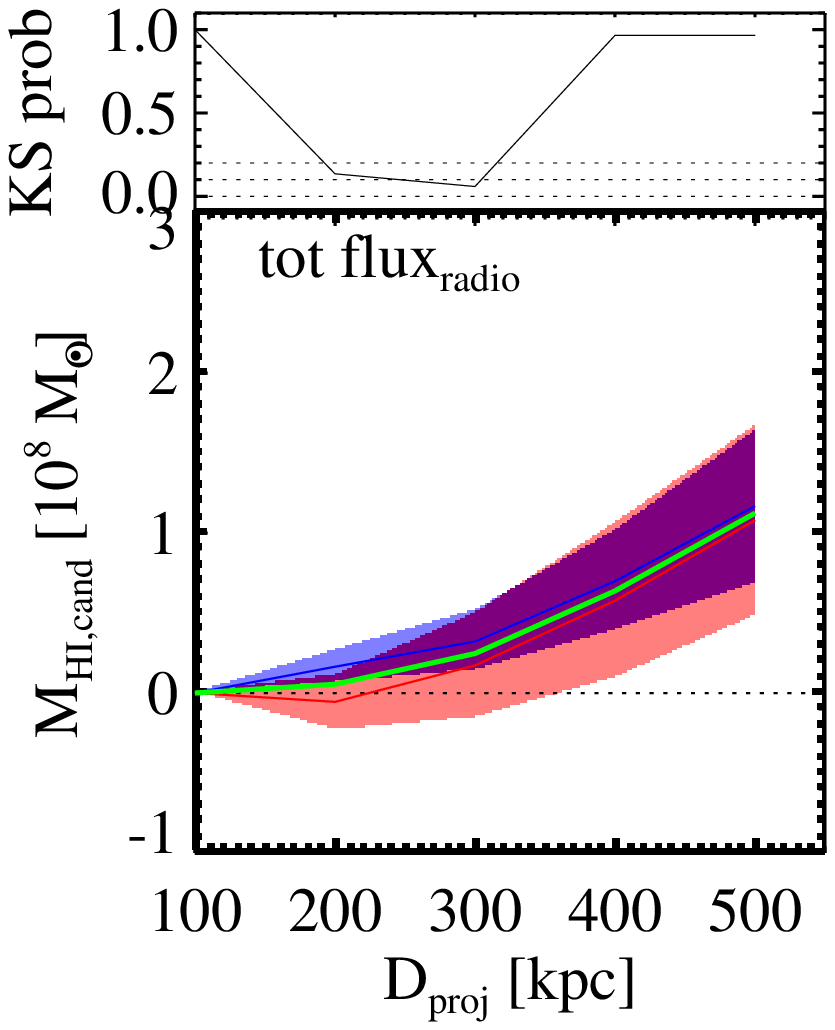}
\hspace{0.5cm}
\includegraphics[width=4.cm]{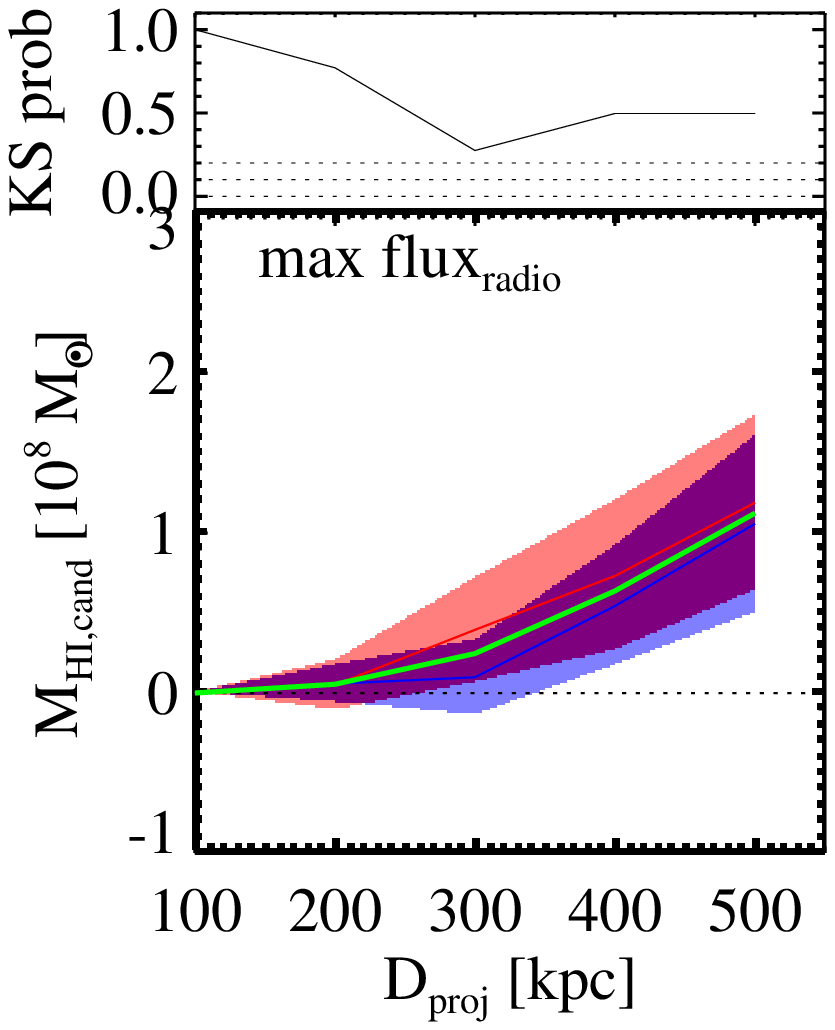}
\hspace{0.5cm}
\includegraphics[width=4.cm]{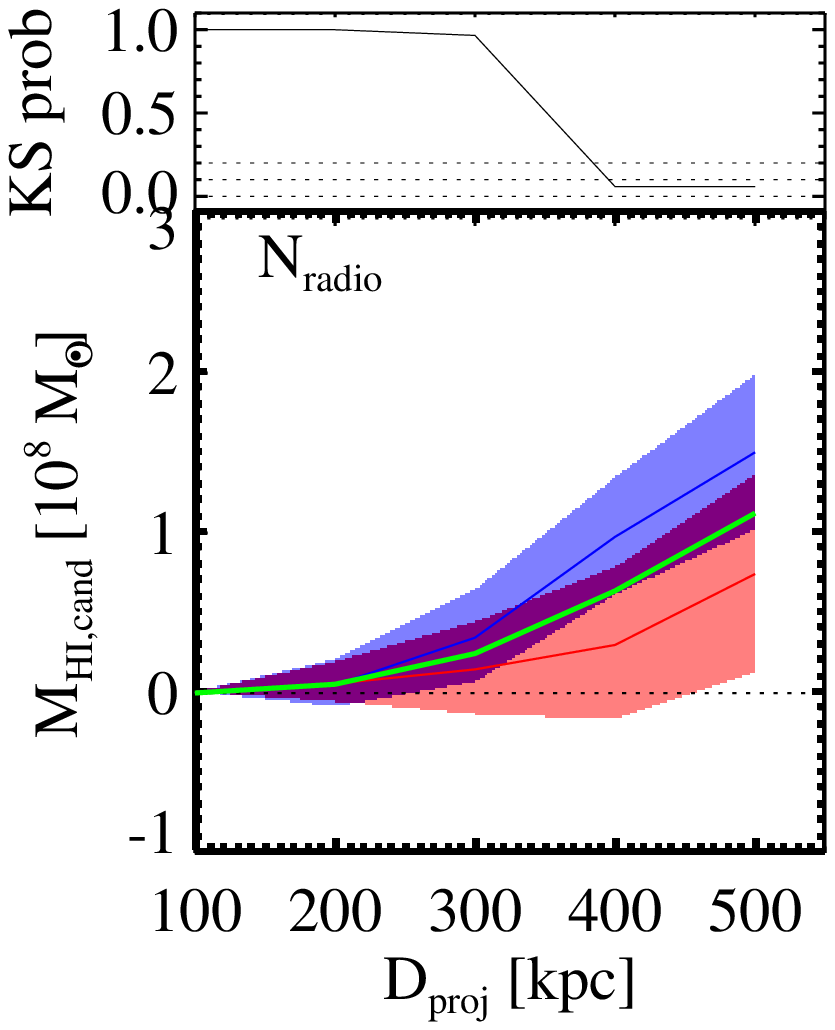}
\caption{Averaged accumulative $\mHIcand$  as a function of distance between 100 and 500 kpc from the primary galaxies.  In each panel, the primary galaxies are evenly divided into two sub-samples by the median value of the parameter denoted in the left-top corner. The red colour is for the sub-sample with higher values of  the dividing parameter, and the blue colour is for the comparison sub-sample with lower values.   The solid lines running through coloured shades show the mean $\mHIcand$ with error bars. The green line shows the averaged $\mHIcand$ for the whole sample of 40 cubes, to guide the eyes.  The black dotted lines show the position of zero \hi mass. The top of each panel shows the KS-test probabilities as a function of radius comparing the distribution of surrounding \hi mass in the two sub-samples. A KS-test probability of 0.1 means that there is 90$\%$ significance that the two distributions are different.
}
 \label{fig:HIprof0}
\end{figure*}

We also find that the flux in candidates is not correlated with radio flux in continuum sources (with a Pearson correlation coefficient of 0.06) or the number of continuum sources (with a Pearson correlation coefficient of -0.27).
We divide the galaxy sample equally into two subsets by the total flux in continuum sources, maximum flux in continuum sources and number of continuum sources, and compare their $\mHIcand$ (see Figure~\ref{fig:HIprof0}). We find no differences for data cubes having different values for the total or maximum radio flux present in the form of continuum sources. There is a trend that data cubes with fewer radio continuum sources have higher $\mHIcand$. However, the number of background continuum sources does not correlate with any of the galaxy properties discussed in section~\ref{sec:conformity}, and we conclude that this correlation does not affect our main results.

Finally, we note that  cumulatively there is a weak gradient of $\mHIcand$ on the low and high V$_{sys}$ sides of the primary galaxies, which is unlikely to be physical but does not  significantly affect the main result of this paper. We refer the readers to Appendix A for a detailed investigation of this issue.

\subsection{A summary of the method}
To summarise, we used SoFiA to extract candidates with typical  mass of 10$^7\ms$, which is lower than that of reliably detected sources but higher than most of the data reduction artefacts. We have shown that the total \hi mass in candidates, $\mHIcand$, is not sensitive to artefacts produced in the continuum subtraction and cleaning processes of our data reduction, and potentially traces a  reservoir of low surface density \hi gas in data cubes. In the following section we study the nature of these candidates and investigate whether $\mHIcand$, i.e., the gas mass contained in the environment of Bluedisk galaxies, correlates with properties of the primary galaxies.

\section{Results}
\label{sec:result}
\subsection{The nature of the HI mass in candidates}
\label{sec:matchopt}
\begin{figure*}
\includegraphics[width=14cm]{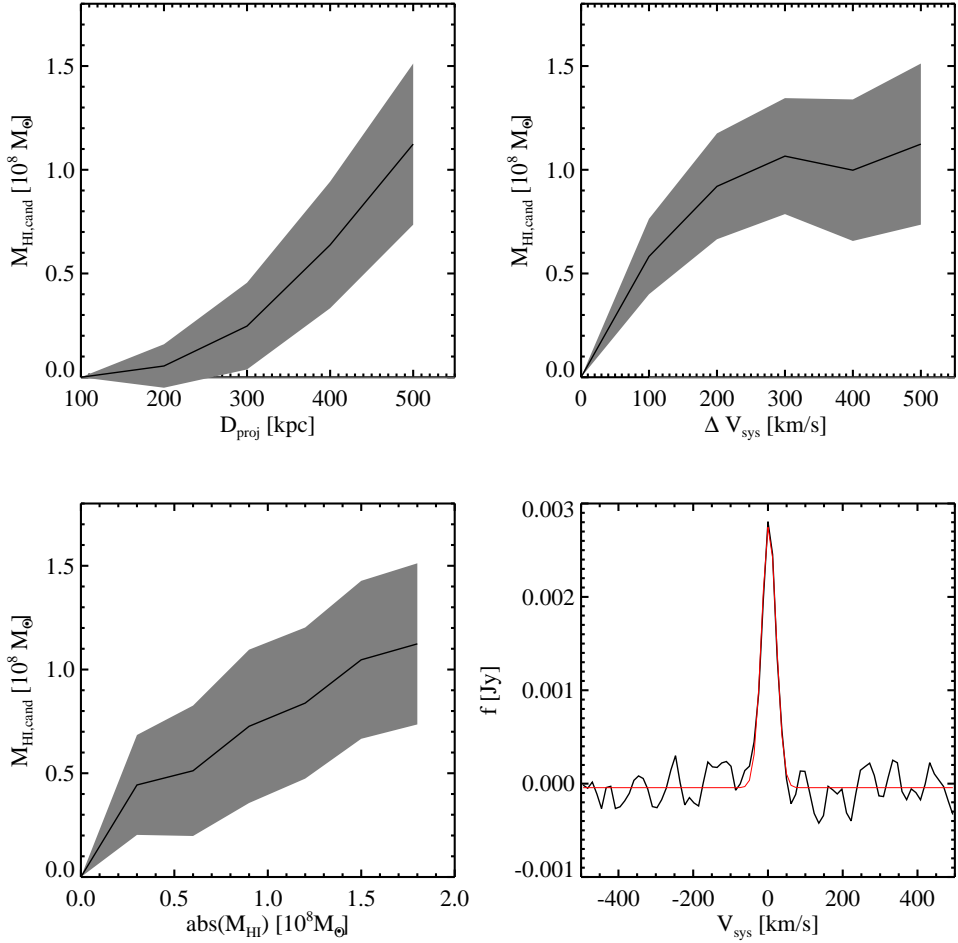}
\caption{Top row and bottom-left panel: the average accumulative $\mHIcand$  as a function of projected distance and systematic velocity distance to the primary galaxies, and as a function of the individual \hi mass (absolute value) of candidates.  The solid black lines show the average and the grey shades show the errors. Bottom-right panel:  stacked spectrum of candidates (black line) and its gaussian fit (red line). }
 \label{fig:mhidistr}
\end{figure*}

In this section,we attempt to understand better the nature of the signal in the cumulative $\mHIcand$ around the Bluedisk primary galaxies.
From the top row of Figure~\ref{fig:mhidistr}, we can see that the average profile of  $\mHIcand$  around the 40 primary galaxies rises steadily to $\sim1.1\times10^8\ms$ from a projected distance of 100 kpc out to 500 kpc,   and from a systematic velocity distance of 0 out to $\sim\pm300$ km$/$s.  
In the bottom-left panel, the accumulative $\mHIcand$ is plotted as a function of absolute \hi mass of individual candidates, and we can see that $\mHIcand$ is not dominated by the most massive candidates. The candidates with $\mHI>10^8\ms$ contributes only $\sim1/3$ of the net signal. We extract the spectra for all the candidates, shift them to have the same central systematic velocity of 7700 km$/$s,  and stack them. The stacked spectrum, as displayed in the bottom-right panel of Figure~\ref{fig:mhidistr}, is a narrow emission line with a gaussian fit $\sigma$ of 18.9$\pm$5.7km$/$s. Removing the resolution effect, the line has a width of $\sigma\sim14.3$ km$/$s.  If we assume rotational systems and apply this velocity width to the baryonic Tully-Fisher relation (McGaugh et al. 2000), it corresponds to a baryonic mass of  4.4$\pm$2.5$\times10^7\ms$, similar to the typical $\mHI$ of a positive candidate.  The candidates could be low mass satellite galaxies, \hi clouds, or \hi clumps in low surface density \hi disks. 

\begin{figure*}
\includegraphics[width=5cm]{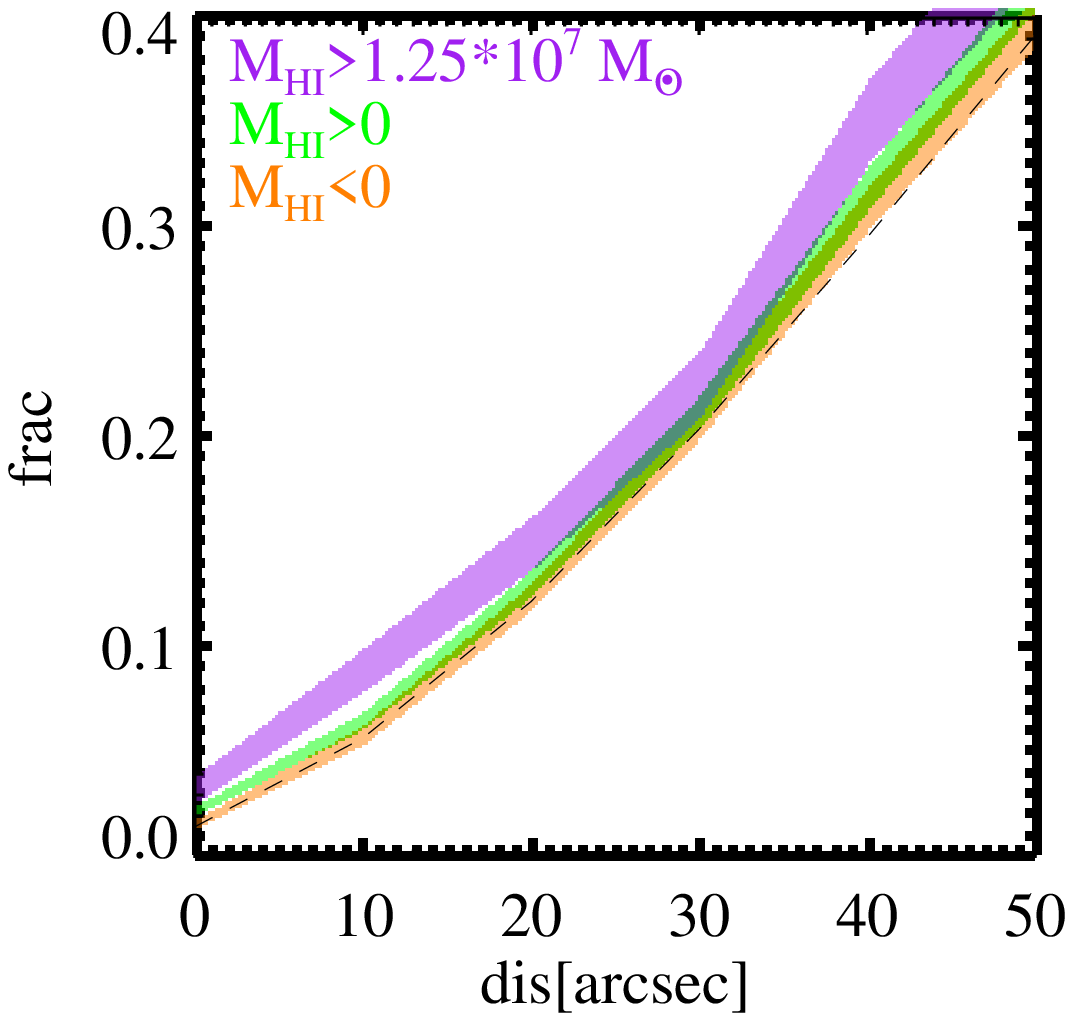}
\hspace{0.2cm}
\includegraphics[width=5cm]{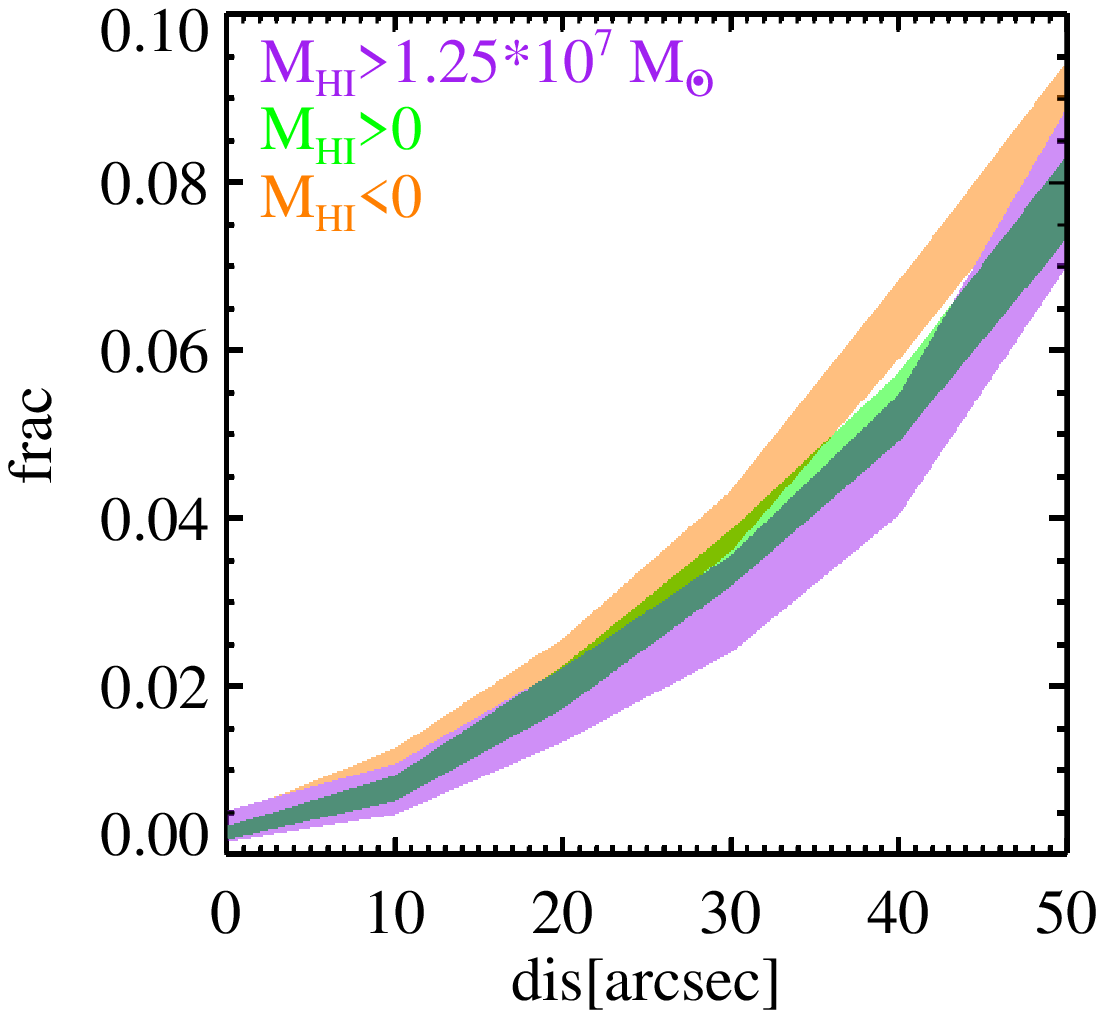}
\hspace{0.2cm}
\includegraphics[width=5cm]{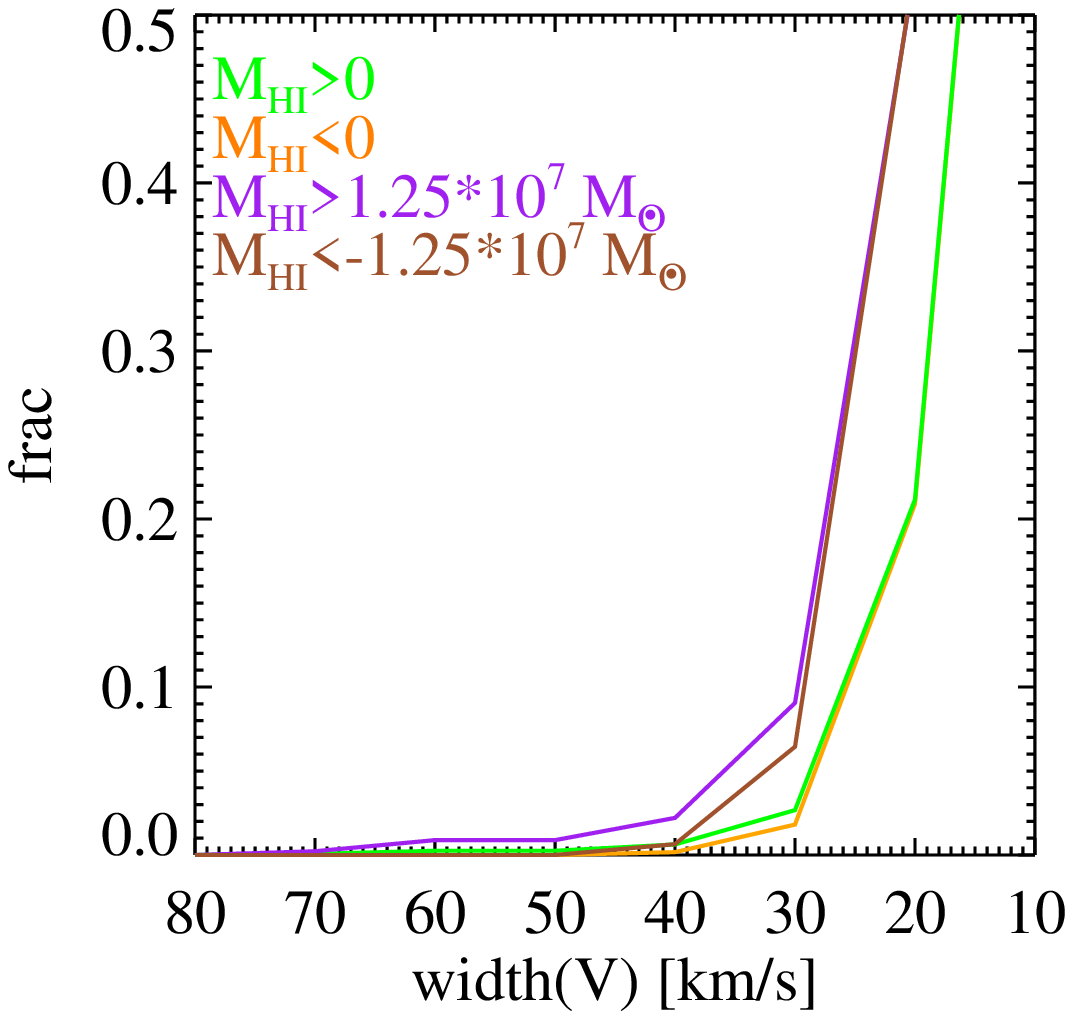}
\hspace{0.2cm}
\caption{The onnection of  negative (orange), positive (green) and bright positive (purple) candidates with optical and radio continuum sources.
 Left panel: accumulative fraction of projected distance between the candidates and their closest optical galaxies. The dashed line is built from matching simulated random positions with the optical galaxies. Middle panel: accumulative fraction of projected distance between the candidates and  radio continuum sources.  Right panel: accumulative fraction of velocity width (length of the detection mask in the velocity direction). The colours used for the different sub-samples are denoted in the left-top corner.}
 \label{fig:match}
\end{figure*}

If part of $\mHIcand$ probes \hi in galaxies, we expect the positive candidates to trace galaxies more closely than the negative candidates. As a test, we 
select galaxies with $r$ band flux brighter than 20 mag from the SDSS DR7 photometric catalog. For each candidate we search for the optical galaxy with the smallest projected distance. We compare the distribution of matching distances for the positive candidates, the negative candidates and simulated random positions.  In the left panel of Figure~\ref{fig:match}, the negative candidates closely follow the curve for random positions (with a K-S test probability of 0.95, meaning that at a 5$\%$ significance  the null hypothesis of the two distributions being drawn from the same parent distribution can be rejected).
We find that a slightly higher fraction of positive candidates can be matched to optical galaxies than the negative candidates at small matching distance  ($<50$ arsec, $\sim$1.5 times the FWHM of the \hi  PSF), although a K-S test probability of 0.28  suggests the difference  is not strong.
If we further take only the positive candidates with $\mHI>1.25\times10^7\ms$ (70 percentile in the mass distribution of positive candidates; based on the bottom-left panel of Figure~\ref{fig:mhidistr}, these are the candidates that contribute nearly all the net $\mHIcand$ signal) into account, we find a significant difference in the distribution of matching distances from the negative candidates (K-S test probability $\sim$ 0.04) and from the  simulated random positions (K-S test probability $\sim$0.01). 
We further check if this small excess of optical matches for the bright positive candidates might be due to residual from the continuum subtraction around continuum sources associated with the optical sources.
  In the middle panel of Figure~\ref{fig:match}, we match the projected position of negative, positive and bright positive candidates to continuum sources. We find the three types of candidates to be indistinguishable in the distribution of matching distances when the distance is smaller than 50 arcsec. The K-S test probability is 0.98 for the comparison between negative and positive candidates. The K-S test probability is 0.14 for the comparison between negative and bright positive candidates, suggesting a weak difference, but the curve shows  that the bright positive candidates are less likely to be found along sight lines of continuum sources than the negative candidates. Hence we can conclude that at least the brightest $\sim1/3$ of the positive candidate sample appears to be connected to galaxies, but not necessarily radio bright galaxies. 

We also expect the positive candidates to have a broader line width than the negative candidates if they are more likely to be associated with galaxies. This expectation is confirmed in the right panel of  Figure~\ref{fig:match}. The K-S test probability is 7$\times10^ {-6}$ for comparison between positive and negative candidates,  and is 0.04 for comparison between the positive and negative candidates with absolute $\mHI>1.25\times10^7 \ms$.

In summary, by measuring $\mHIcand$, we add up the signal from unresolved, low \hi mass objects distributed in a large volume ($\sim$500 kpc) around the primary galaxies and other reliably detected objects. It is very likely that at least a fraction of these candidates are small galaxies. The remainder might be galaxies too (e.g., objects below the 20 mag cut in r band used here), or small \hi clouds in the circumgalactic medium.

\subsection{The HI mass in candidates and the properties of primary galaxies}
\label{sec:conformity}
 In this section, we compare $\mHIcand$ between galaxies with different properties, especially between galaxies that are rich in \hi and normal galaxies. We assume that the ${\rm H}{\textsc i}$-rich (either with high $\fHI$ or \hi excess) galaxies have been accreting gas, and hope to find clues about gas accretion by studying $\mHIcand$ around the galaxies.  

\begin{figure*}
\includegraphics[width=3.8cm]{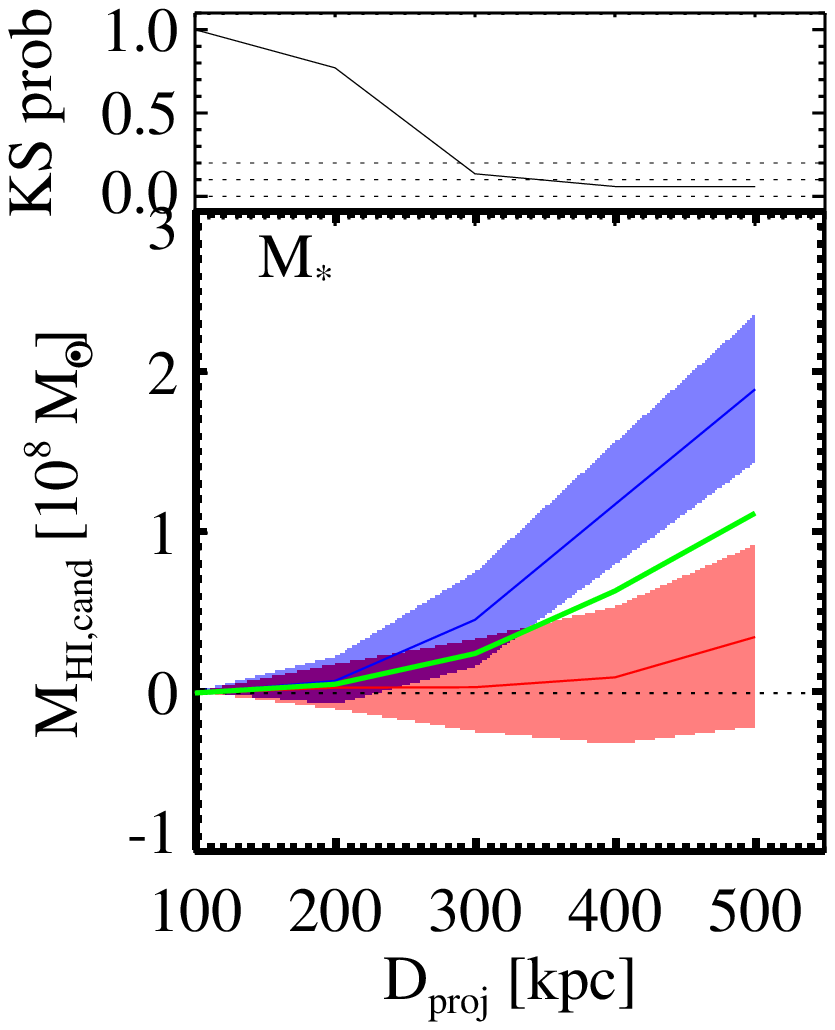}
\hspace{0.2cm}
\includegraphics[width=3.8cm]{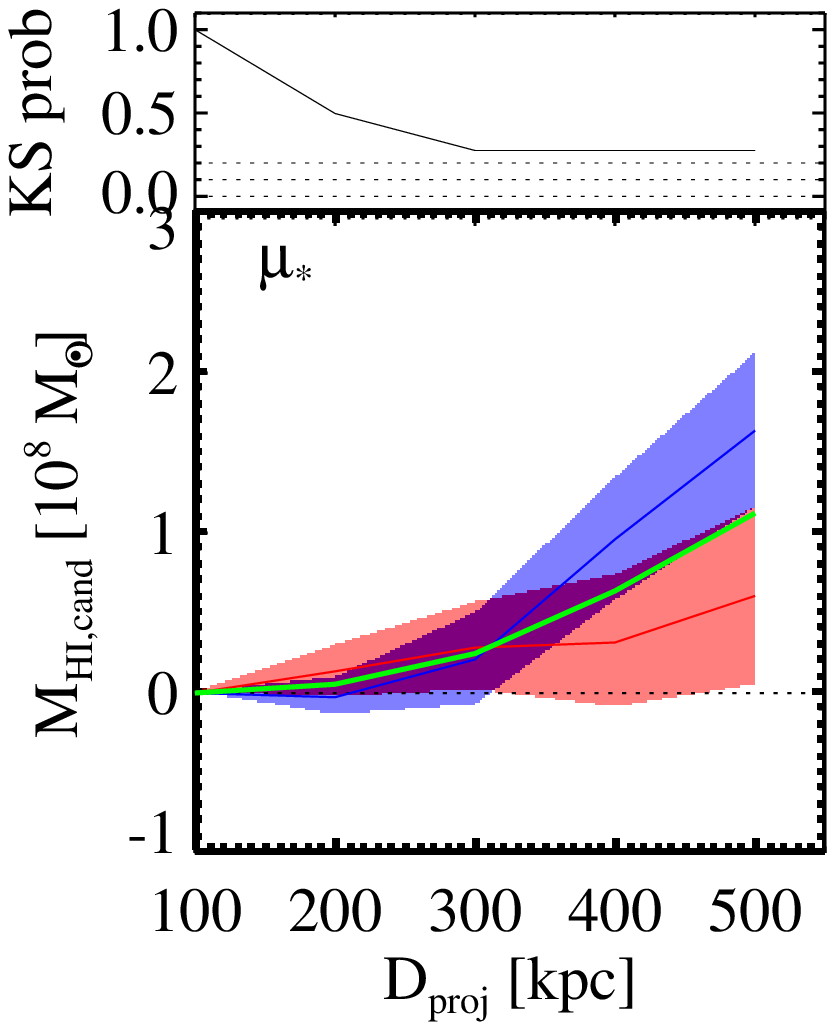}
\hspace{0.2cm}
\includegraphics[width=3.8cm]{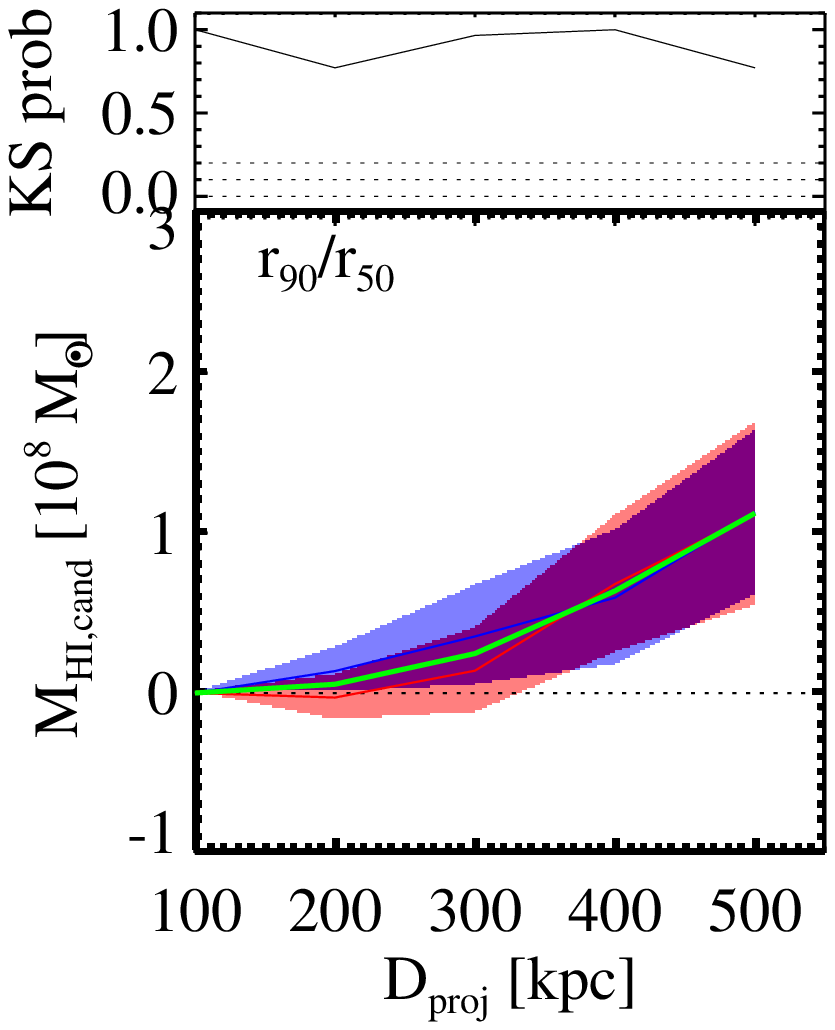}
\hspace{0.2cm}
\includegraphics[width=3.8cm]{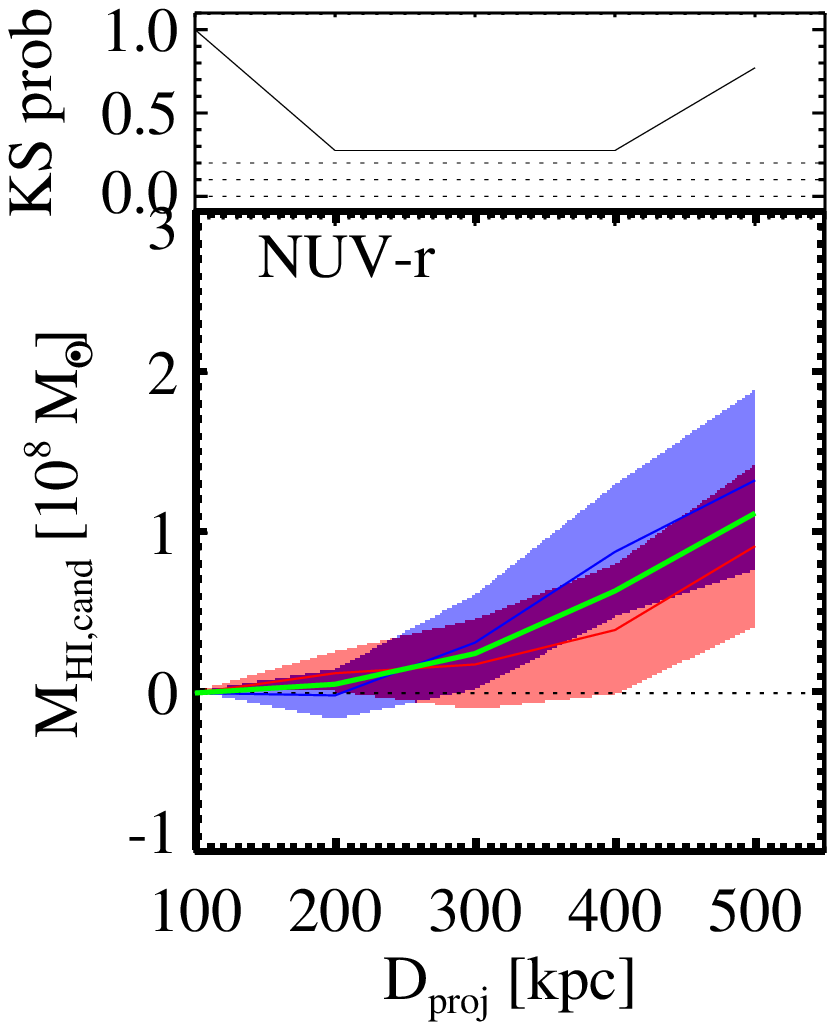}

\vspace{0.5cm}
\includegraphics[width=3.8cm]{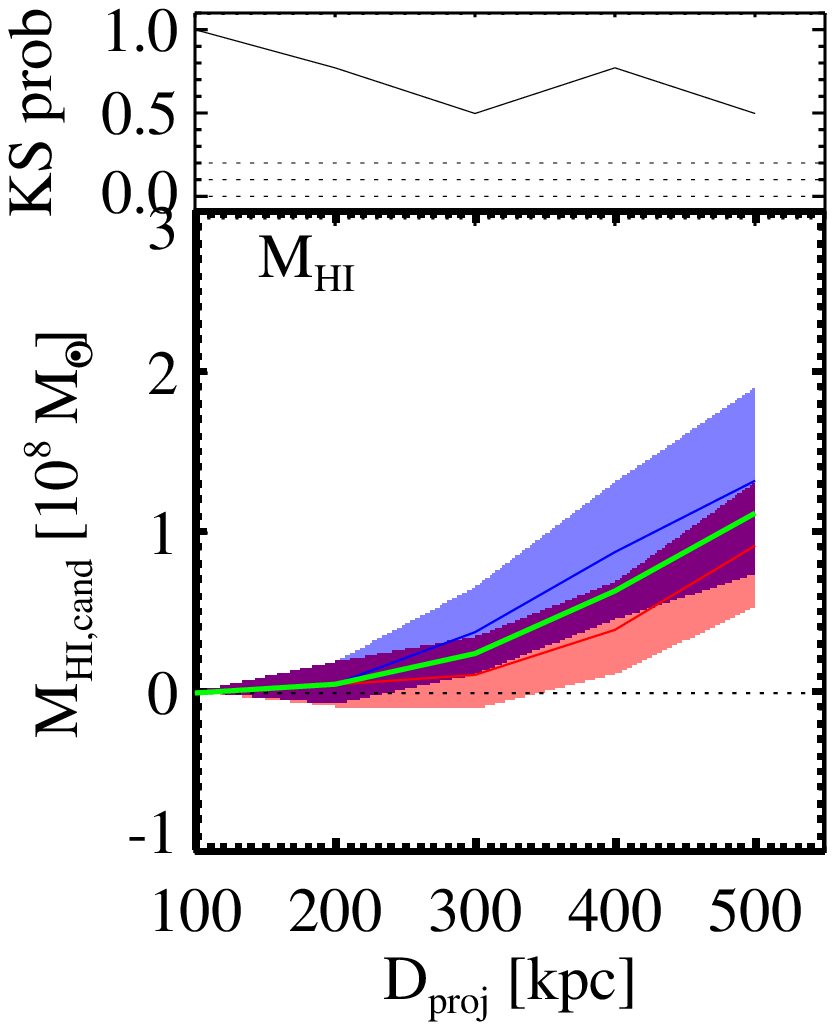}
\hspace{0.2cm}
\includegraphics[width=3.8cm]{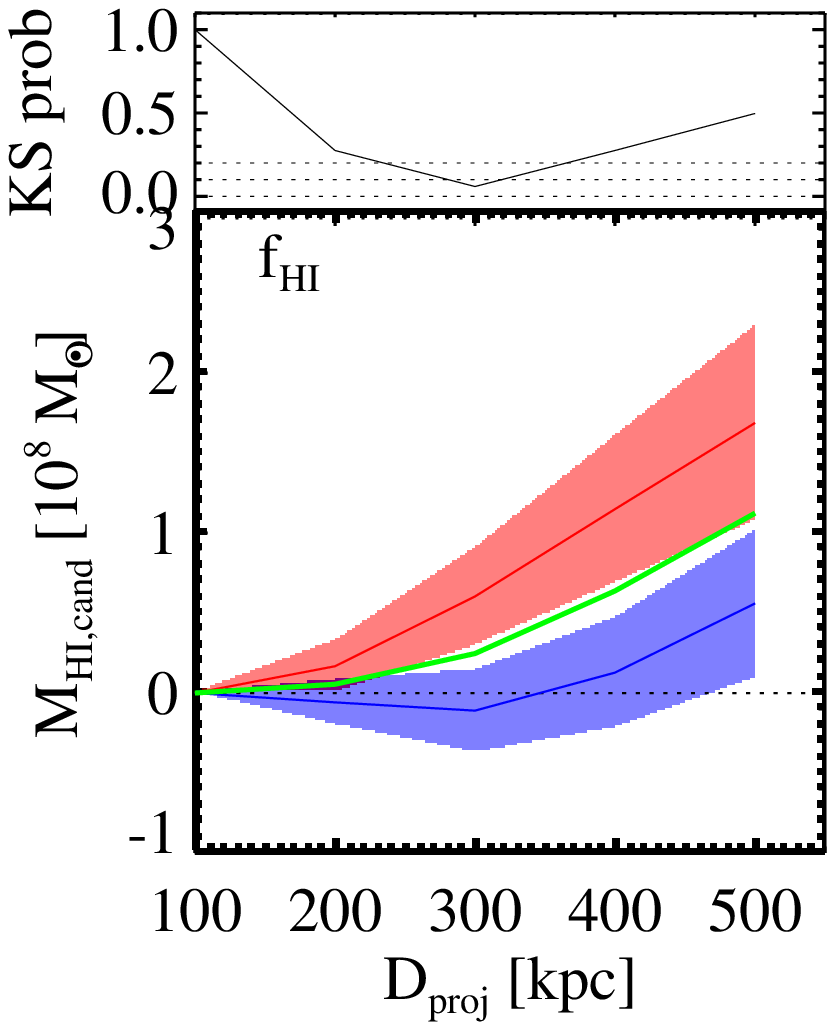}
\hspace{0,2cm}
\includegraphics[width=3.8cm]{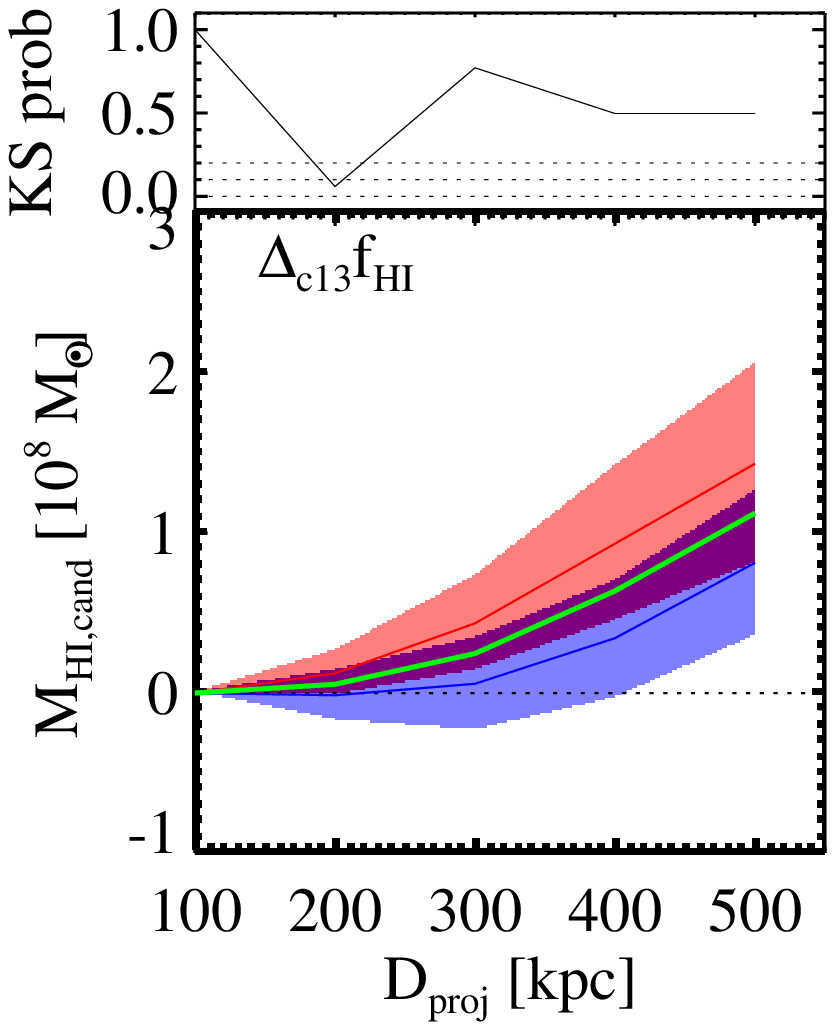}
\hspace{0,2cm}
\includegraphics[width=3.8cm]{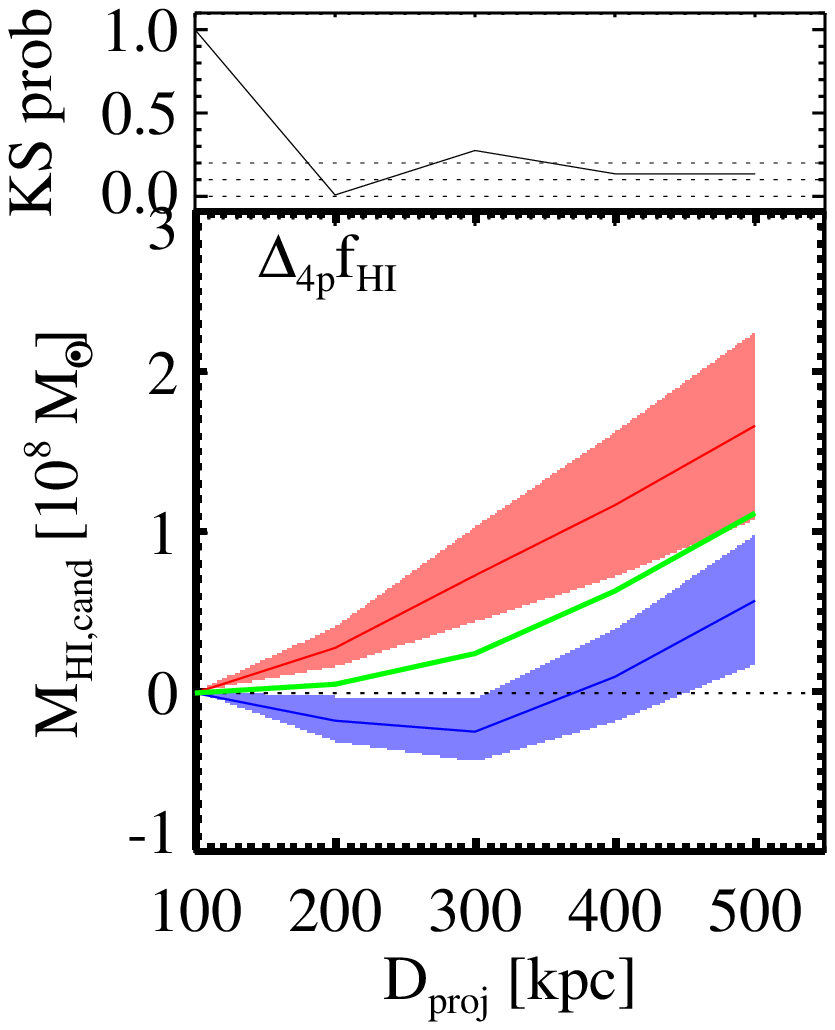}
\caption{Averaged accumulative $\mHIcand$  as a function of distance between 100 and 500 kpc from the primary galaxies.  In each panel, the primary galaxies are evenly divided into two sub-samples by the median value of the parameter denoted in the left-top corner. The red colour is for the sub-sample with higher values of  the dividing parameter, and the blue colour is for the comparison sub-sample with lower values.   The solid lines running through coloured shades show the mean $\mHIcand$ with error bars.   The green solid lines show the averaged accumulative $\mHIcand$ of the whole sample for guiding the eye. The black dotted lines show the position of zero \hi mass. The top of each panel shows the KS-test probabilities as a function of radius comparing the distribution of surrounding \hi mass in the two sub-samples.  The dotted lines show the positions of 0, 0.1 and 0.2.  A KS-test probability of 0.1 means that there is 90$\%$ significance that the two distributions are different.
}
 \label{fig:HIprof}
\end{figure*}

In Figure~\ref{fig:HIprof}, we divide the sample of primary galaxies evenly into two sub-samples, focusing on different properties, and investigate their difference in averaged accumulative $\mHIcand$ as a function of $D_{proj}$ between 100 and 500 kpc. The dividing galactic properties include the stellar mass (M$_*$), the stellar mass surface density ($\mu_*$), optical concentration (r$_{90}/$r$_{50}$), NUV$-$r colour, the \hi mass ($\mHI$), \hi mass fraction ($\fHI$), and \hi excess ($\Delta_{C13}\fHI$ and $\Delta_{4p}\fHI$) \footnote{We point out the pairs of subsamples investigated here all have similar ranges of redshift (e.g. the difference in median luminosity distance is always less than 6\%)}.   

The most prominent trend is  the sub-sample with lower M$_*$  having  higher $\mHIcand$ than the comparison sub-sample, with a higher than 90\% significance that the two distributions are different at a $D_{proj}$ of 500 kpc.  
The sub-samples with higher $\fHI$ and $\Delta_{4p}\fHI$ also have considerable higher $\mHIcand$ than the comparison sub-samples; especially the latter has a close to 90\% significance for the difference in distribution from the comparison sample. 
There are also weak  trends that the sub-sample with lower $\mu_*$ and bluer NUV$-$r colour  have higher $\mHIcand$ (considering the difference in average values and the close-to 0.2 K-S test probabilities).  There is no significant correlation between $\mHIcand$ and the optical concentration, the \hi mass or  $\Delta_{C13}\fHI$ of primary galaxies. 

\begin{figure*}
\includegraphics[width=17.cm]{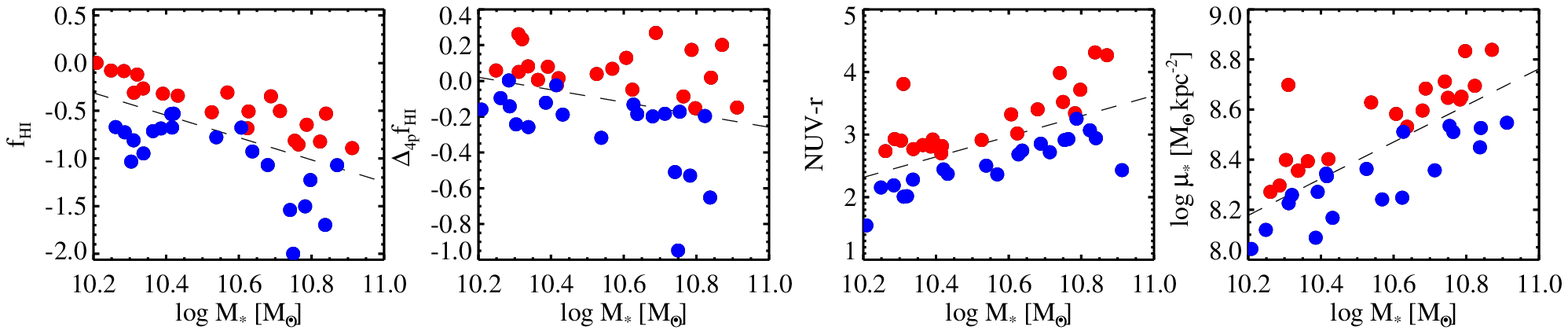}
\vspace{1cm}

\includegraphics[width=3.8cm]{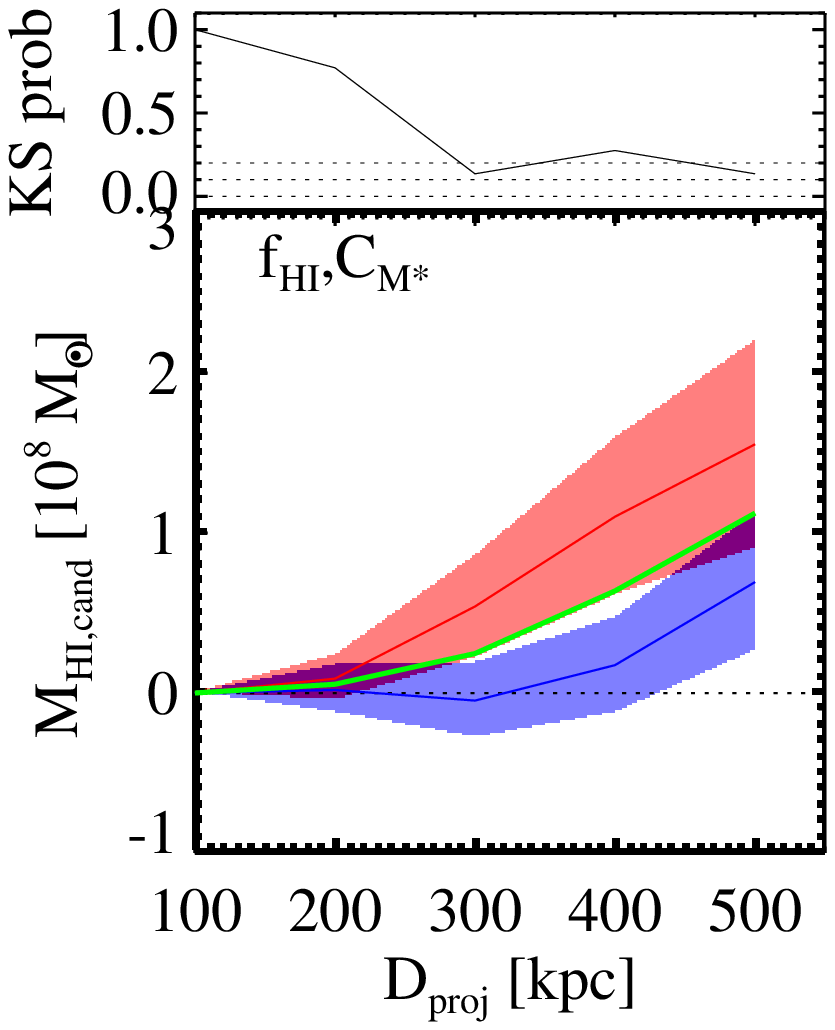}
\hspace{0.2cm}
\includegraphics[width=3.8cm]{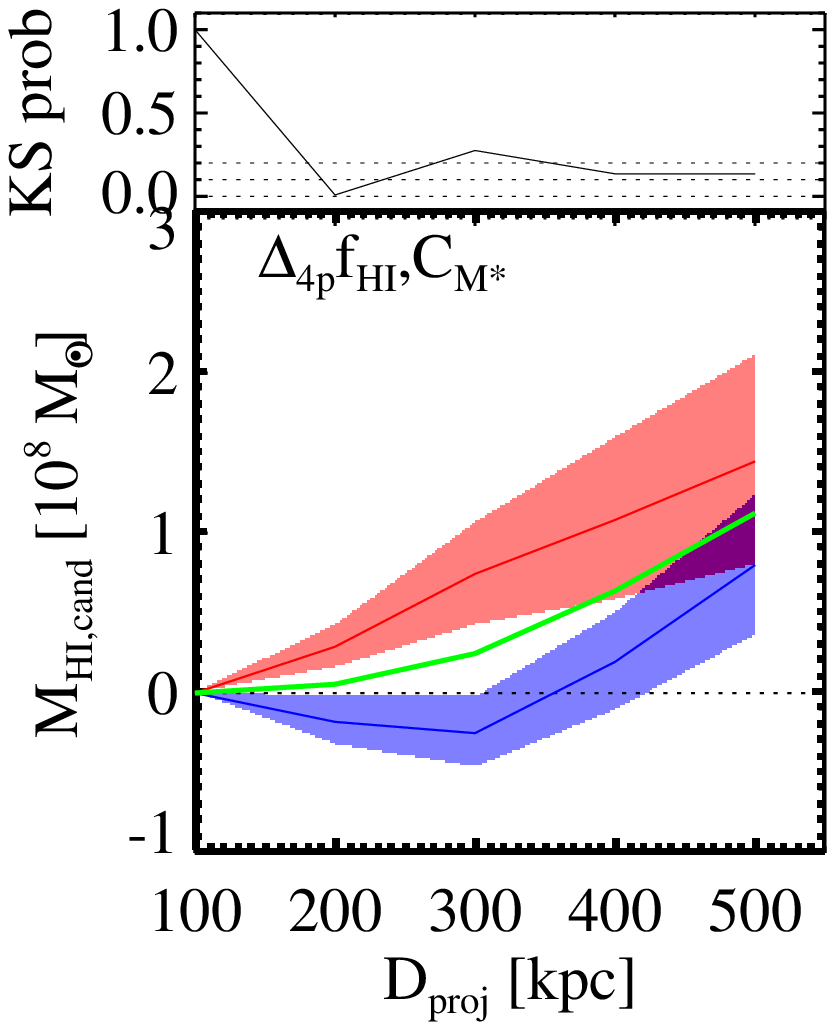}
\hspace{0.2cm}
\includegraphics[width=3.8cm]{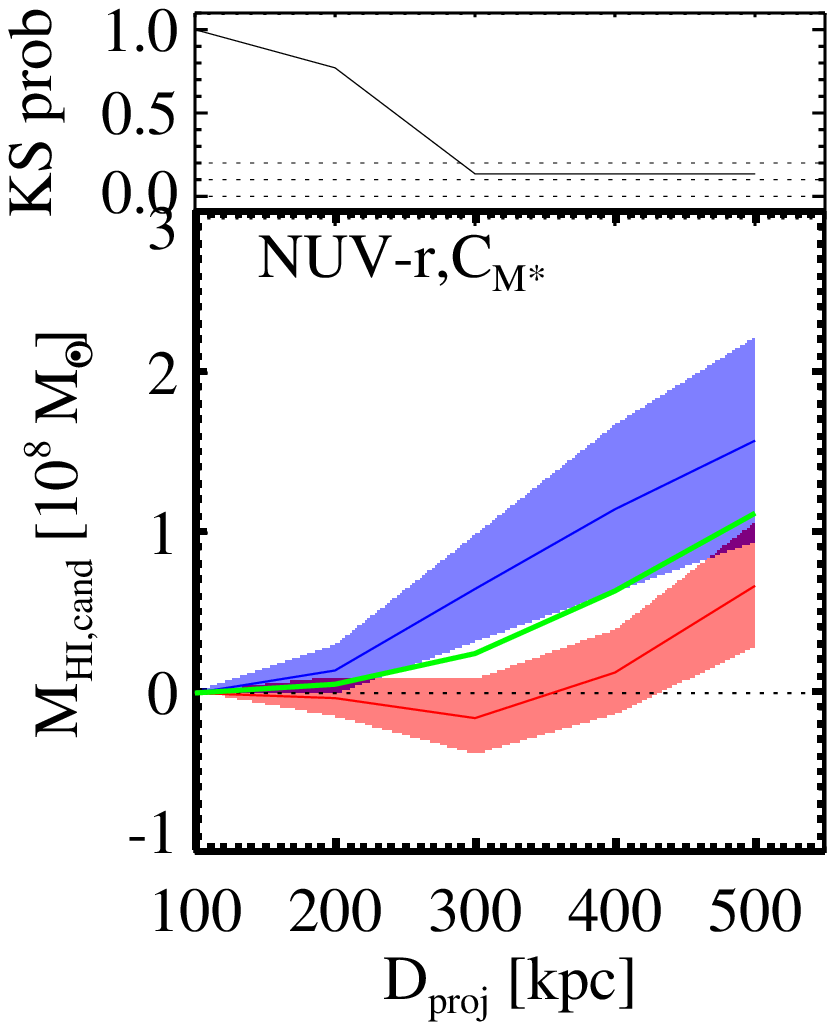}
\hspace{0.2cm}
\includegraphics[width=3.8cm]{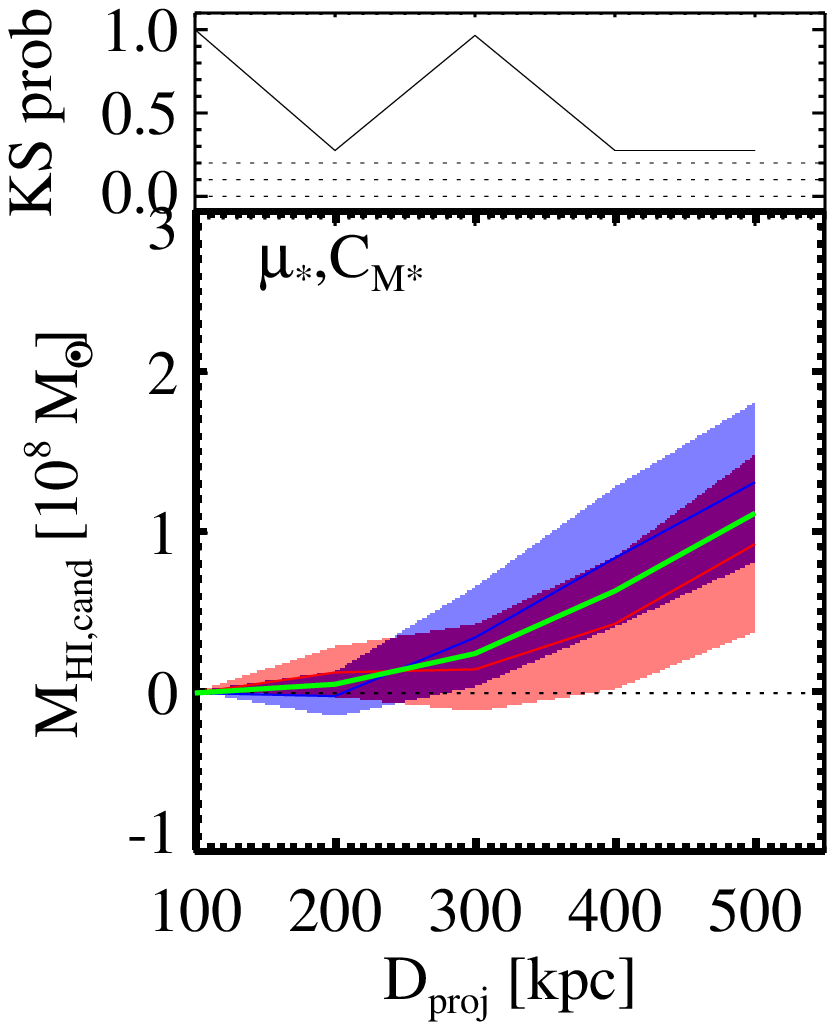}
\caption{The top row demonstrate how the sample is divided with stellar mass controlled (see text). In each panel, the dashed line is the linear fit to the whole sample, and the red and blue dots are with vertical distance above and below the median to the dashed line.  The bottom row is similar as Figure~\ref{fig:HIprof}, but the sub-samples are divided with stellar mass controlled as shown in the panel on top.
}
\label{fig:HIprof2}
\end{figure*}

We further control stellar mass to reveal a more intrinsic correlation between $\mHIcand$ and  \hi richness  of  primary galaxies. For each of the sample dividing parameters, we derive the linear relation with stellar mass,  and calculate vertical distance  to the relation. We use this distance to divide the primary galaxies evenly into two samples, and study their difference in averaged accumulative $\mHIcand$. In Figure~\ref{fig:HIprof2}, we only consider the dividing parameters which have already shown an indication for a correlation with $\mHIcand$ in Figure~\ref{fig:HIprof}.
We can see the trend of galaxies with higher $\fHI$ having on average higher $\mHIcand$ becomes more prominent as suggested by the K-S test probabilities. The trend as a function of $\Delta_{4p}\fHI$ remains almost unchanged, suggesting the advantage of $\Delta_{4p}\fHI$ as an unbiased measure of \hi richness. 
  Bluer primary galaxies also have on average higher $\mHIcand$.  $\mHIcand$ is no longer dependent on $\mu_*$ of primary galaxies.  These trends are consistent with each other, for at the same stellar mass, ${\rm H}{\textsc i}$-rich galaxies are also more star forming galaxies. Finally, we note that, when the sample is divided by $\fHI$, $\Delta_{4p}\fHI$  and NUV$-$r of the primary galaxies, the difference in the  $\mHIcand$ distribution of the sub-samples is tentative, with a K-S test probability of slightly above 0.1 at a $D_{proj}$ of 300-500 kpc. 

\section{Discussion: evidence of gas conformity}
\label{sec:discussion}
We have found evidence for the presence of an excess of faint gas clouds in a volume out to 500 kpc in projected distance and 500 km$/$s in velocity around \hi excess galaxies, i.e., galaxies that have more \hi mass than other galaxies with similar optical properties. 
A parallel study of individual satellite galaxies around the Bluedisk primary galaxies revealed that satellites around \hi excess galaxies have higher \hi excess than satellites around galaxies with normal \hi contents (E. Wang et al. 2015). Although these two studies are based on different signal-to-noise levels of the \hi data, reflecting different reservoirs of cold gas,
they consistently tell the same story: the \hi reservoir in the environment of (massive) galaxies appears to reflect the \hi content of those galaxies.
As discussed in K10, this environmental \hi may trace the dense gas tunnelled into halos through filaments, the so-called ``cold mode'' accretion (Kere{\v s}  et al. 2005). 

Recent numerical studies have shown that  cold mode accretion  is not affected by galactic winds (Powell et al. 2011), hence it can easily leave its footprint on the low-mass satellite galaxies. 
Beside the gas locked into satellites,  our \hi candidates may also include the \hi clouds that condense from the ionised cold mode gas (Kere{\v s} et al. 2009a, 2009b). 
It is interesting to note that the lower limit in the \hi mass of candidates detected in our data ($\sim10^6\ms$, imposed by our 4-$\sigma$ detection threshold) is roughly the lower mass limit required by an \hi cloud to survive in a hot gas halo around galaxies (Murray \&Lin 2004).
Due to the lack of baseline spacings short-ward of 36 m, we are unable to detect the very extended, smoothly distributed gas. However, the detected candidates that trace the peaks in the IGM or small companion galaxies, indicate that an underlying reservoir of gas from cold mode accretion might be present. 

In the picture of cold mode accretion, the outer region of  the galactic disks is the favoured location of the accretion taking place (as in the classical model of hot accretion) (Pichon et al. 2011). The accretion rate is bursty because the inflow filament is clumpy (Brooks et al. 2009). These two factors may produce the excess \hi in ${\rm H}{\textsc i}$-excess galaxies.
The Bluedisk primary galaxies have stellar masses around that of the MW, for which around 50$\%$ of the gas accretion occurs in the cold mode (van de Voort \& Schaye 2012). 
The most prominent trend found in this paper is that the primary galaxies with lower stellar masses have higher  \hi mass in candidates around them, which may suggest a transition between cold- and hot-mode accretion, as predicted by the simulations (Dekel \& Birnboim et al. 2006).
Recently  Kauffmann (2015) found a tendency for the satellites to align along the major axis of late-type ${\rm H}{\textsc i}$-rich central galaxies with M$_*<10^{10.5} \ms$; they found no such phenomenon for more massive central galaxies. Their result also suggests such a transition. The median M$_*$ of our primary galaxies ($10^{10.56} \ms$ ) is consistent with their suggested transition point.

An alternative picture  is that the accumulative signal from \hi candidates traces a rather smooth accretion of dark galaxies.
Genel et al. (2010) studied the growth of dark matter halos and found that at least $40\%$ of the baryons in the halos are accreted from very low mass halos, containing smooth cold $T\sim10^4$K gas and no stars. Although no dark galaxies have been found so far from nearby surveys (Zwaan\& Briggs 2000, Zwaan 2001, Koribalski et al. 2004, Haynes 2007), this can be explained if dark galaxies are preferentially found around gas accreting massive galaxies: galaxies with very high $\fHI$ and as massive as the Bluedisk sample are rare in the nearby universe. 
This picture may still be consistent with the cold accretion scenario, since recent simulations show that halos in filaments may have more sub haloes than those in other environment (Guo et al. 2014). 

To this date no gas accreting filaments or clouds have directly been observed to exist within a few tens of kpc around galaxies in the local universe, even when the observational depth reaches 10$^{19}$ atom cm$^{-2}$, as in the case of the WSRT Hydrogen Accretion in LOcal GAlaxies Survey  (HALOGAS, Heald et al. 2011). However, numerical simulations demonstrate that the accreting gas may just get slowed down and  ionised through interaction with the hot gas halo when it gets close to the galactic disks (Putman et al. 2012).  In our study here, the candidates are beyond 100 kpc from the galactic centre, far above the disk-halo interface.

Finally, we address the question if  the merging of satellites contributes significantly to the cold gas in primary galaxies. We select the 20 primary galaxies with higher than median $\fHI$, and only 8 of them have reliably detected satellites within a $D_{proj}$ of 500 kpc and $\Delta V_{sys}$ of 500 km$/$s. 
 We calculate the timescale for a pair of galaxies to merge following the empirical formula from Kitzbichler \& White (2008), 
 \begin{equation}
  T=1.6\frac{r}{25h^{-1}kpc} (\frac{M_*}{3\times10^{10}h^{-1}\ms})^{-0.3} Gyr  
  \end{equation}
  where M$_*$ is the total stellar mass of the pair.
  We replace M$_*$ in the formula with $\mHI$ plus M$_*$ to minimize the timescale.
We calculate the gas merging rate, and compare it with the SFR  for these 8 primary galaxies with satellites. The highest ratio is 0.12, hence the merging rate is too low to support the SFR, and can not be a major source for replenishing the gas storage in the Bluedisk ${\rm H}{\textsc i}$-rich galaxies.
It is consistent with the findings of many previous studies like K10 and Di Teodoro \&Fraternali (2014). 

The  gas-rich satellites studied by K10 and E. Wang et al. (2015) might be the  tip of the iceberg of the missing gas accretion. Here with the detection of an excess in the low-mass low-column density \hi environment, we might have detected a substantial part of the iceberg below the sea-level.

\section{Summary and future prospects}
We have developed a new technique  to extract information about the large-scale distribution of \hi mass  from 21 cm synthesis data cubes through a technique similar to stacking, which works in an  \hi mass regime which is lower than achievable for reliable direct detections, but still high enough not to be dominated by data reduction artefacts.. 
We used our technique to investigate the mass contained in candidate \hi sources, possibly associated with individual satellite galaxies, around the Bluedisk primary galaxies. We found a connection between \hi excess ($\Delta_{4p}\fHI$, or high values of $\fHI$ at a fixed stellar mass) in primary galaxies and excess of \hi in low mass candidates in the surroundings.  These \hi masses may be the  tip of the iceberg of an underlying extended reservoir of gas that fuels the primary galaxies.
 It is a direct detection of the galactic gas conformity phenomenon down to very low \hi masses ($\sim10^7 \ms$). 
The result is consistent with the cold mode accretion in cosmological simulations. In such a picture, primary galaxies and satellites are both fuelled by the cosmic web and exhibit the conformity phenomenon of gas richness. The unusually blue outer discs in Bluedisk galaxies with high \hi mass fractions, indicating inside-out formation, can also be explained with such cosmological gas accretion (Kauffmann 1996).

At this moment, we are unable to establish  general relations between the mass, structure and morphology of the stellar disks and the surrounding \hi masses because we are limited by selection effects in our sample of primary galaxies. Our sample misses the most massive galaxies (with M$_*>10^{11}\ms$) in the universe. In more massive halos, cold gas around primary galaxies  may behave in a different way from what we find here, for those galaxies are more strongly affected  by the halo (Catinella et al. 2014). 
Limited by primary beam effects, we are also unable to investigate how far away from the primary galaxies the conformity phenomenon extends.
Limited by the small sample size,  we are unable to statistically measure the total \hi mass in all the satellite galaxies.  Because the accumulative signal ($\mHIcand$) we obtain is only $\sim$3 $\sigma$ above 0, we are unable to calculate spatial gradients, to quantify spatial distribution, or to estimate inflow rate. 
It  also remains unclear how  the accreting gas arrives at the ${\rm H}{\textsc i}$-excess galaxies, and how much of the \hi mass in candidates is not locked into optical galaxies (i.e. in the form of clumps in the more diffuse regions in and between cosmic filaments, or in dark galaxies).


We note that the emphasis of this paper is the exploration of  a new technique. We demonstrate the promising potential to investigate \hi properties  in the radio synthesis cubes below 5-6 $\sigma$, which is generally considered as a threshold for detection reliability.  We have tried to control the systematics as best as possible and reach the tantalising result that we may have detected evidence for the presence of excess \hi in the intergalactic medium around \hi rich galaxies.
We look forward to applying the technique and analysis presented in this paper to a much larger, more uniform and complete dataset, especially from the ASKAP \hi All-Sky Survey, known as WALLABY, and the proposed WSRT Northern Sky \hi Survey (WNSHS), both described in Koribalski (2012).

\section*{Acknowledgements}
We thank the anonymous referee for constructive comments. We gratefully thank T. Oosterloo, V. Kilborn, B. Sault, L. Staveley-Smith and S. Huang for useful discussions. 
J.M. van der Hulst acknowledges support from the European Research Council under the European Union's Seventh Framework Programme (FP/2007-2013) / ERC Grant Agreement nr. 291531.
J. Fu acknowledges the support from the National Science Foundation of China No. 11173044 and the Shanghai Committee of Science and Technology grant No. 12ZR1452700.
T. Xiao acknowledges the support from NSFC under Grant No. 11203056.

GALEX (Galaxy Evolution Explorer) is a NASA Small Explorer, launched in April 2003, developed in cooperation with the Centre National d'\'{E}tudes Spatiales of France and the Korean Ministry of Science and Technology.

Funding for the SDSS and SDSS-II has been provided by the Alfred P. Sloan Foundation, the Participating Institutions, the National Science Foundation, the U.S. Department of Energy, the National Aeronautics and Space Administration, the Japanese Monbukagakusho, the Max Planck Society, and the Higher Education Funding Council for England. The SDSS Web Site is http://www.sdss.org/.

This publication makes use of data products from the Wide-field
Infrared Survey Explorer, which is a joint project of the University
of California, Los Angeles, and the Jet Propulsion Laboratory/California
Institute of Technology, funded by the National
Aeronautics and Space Administration.

\bibliographystyle{mn2e}

\appendix
\section{$\mHIcand$ on the low and high V$_{sys}$ sides of the primary galaxies}
\begin{figure*}
 \includegraphics[width=3.cm]{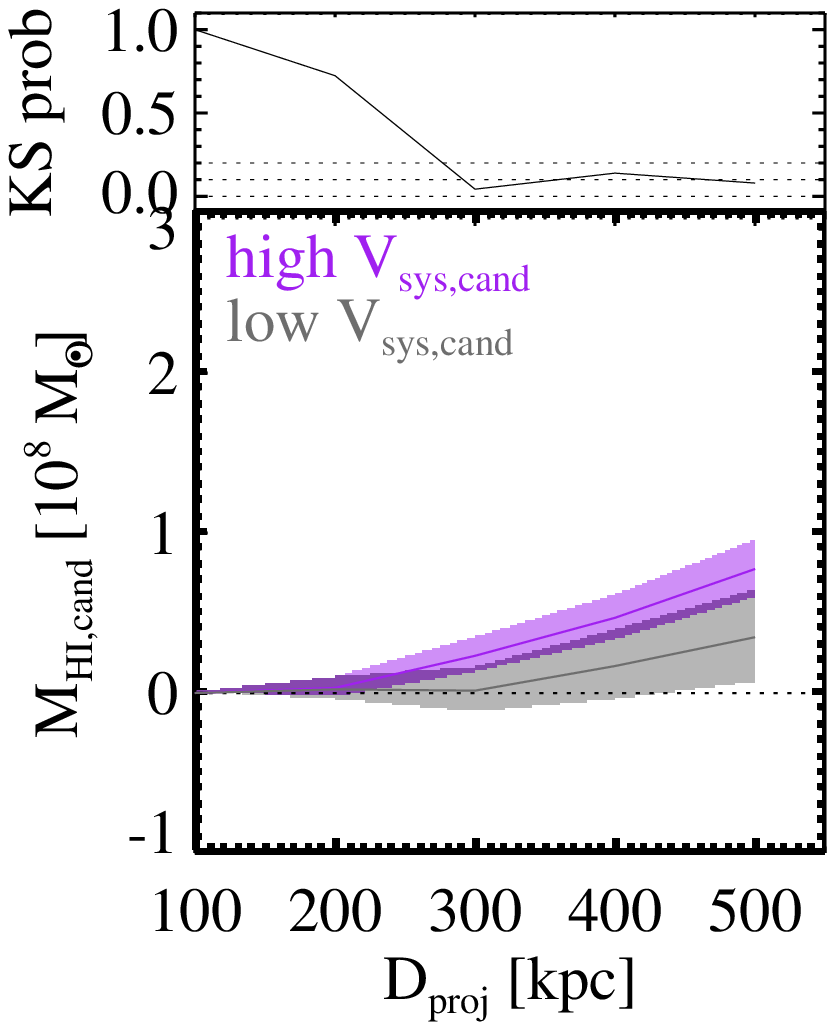}
\hspace{0.2cm}
\includegraphics[width=3.cm]{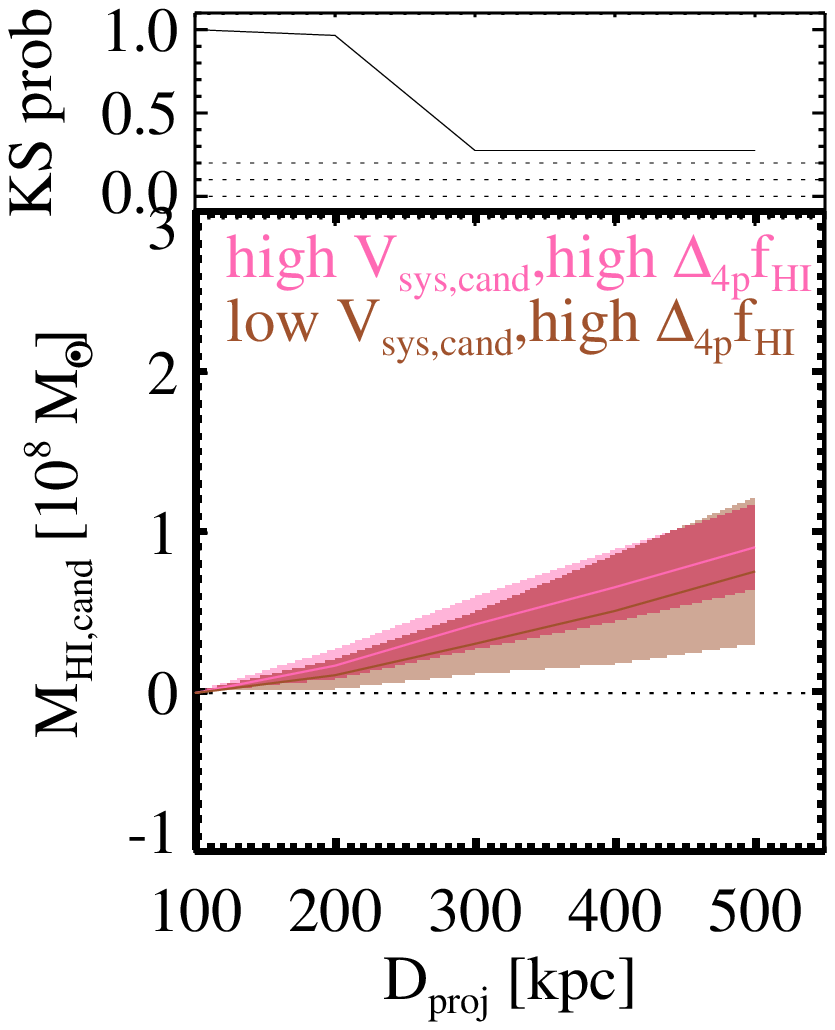}
\hspace{0.2cm}
\includegraphics[width=3.cm]{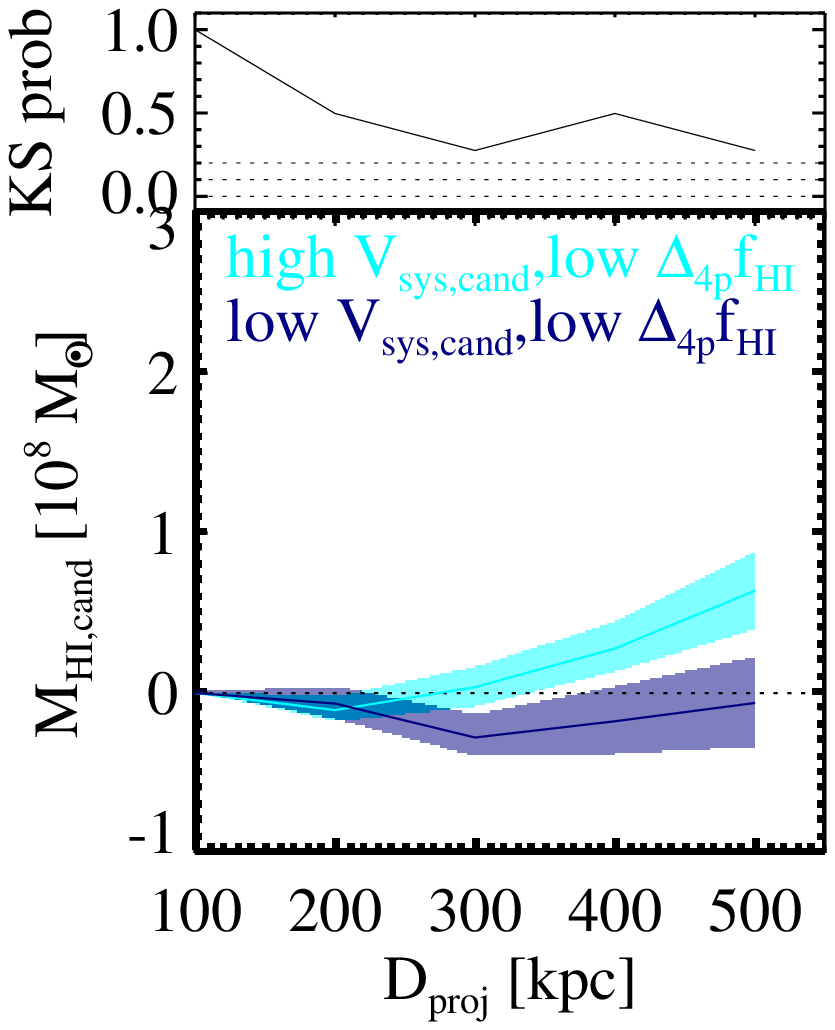}
\hspace{0.2cm}
\includegraphics[width=3cm]{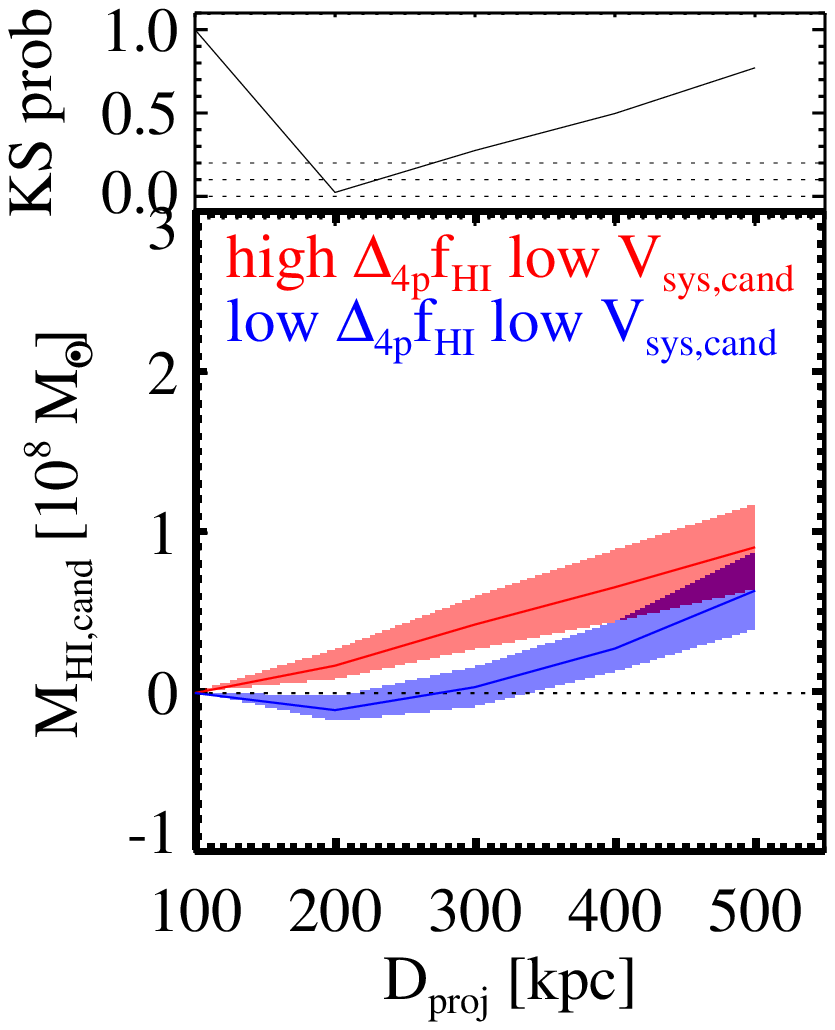}
\hspace{0.2cm}
\includegraphics[width=3.cm]{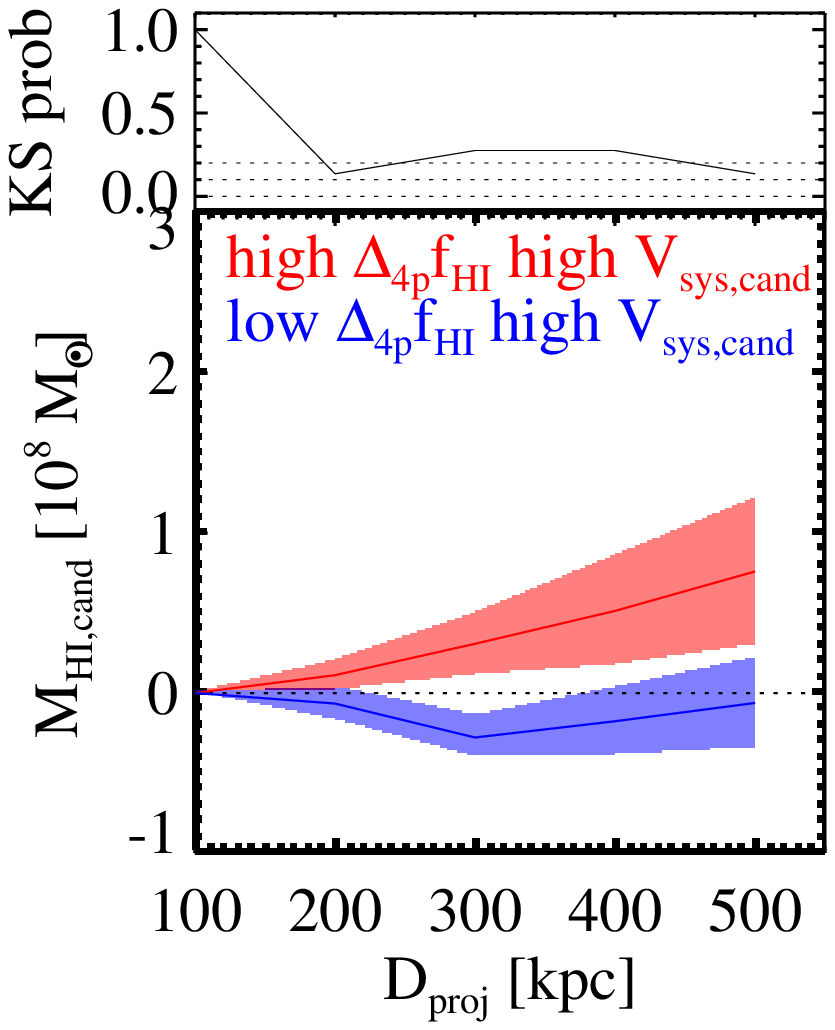}

\vspace{0.4cm}

 \includegraphics[width=3.cm]{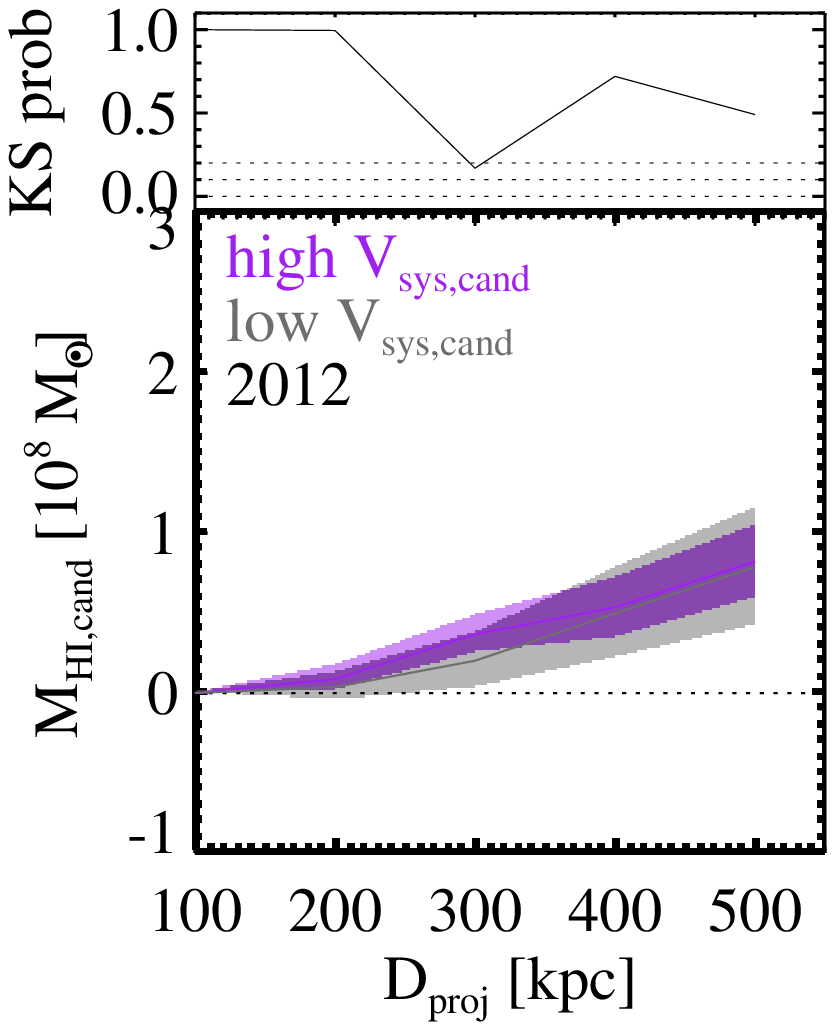}
 \hspace{0.2cm}
\includegraphics[width=3.cm]{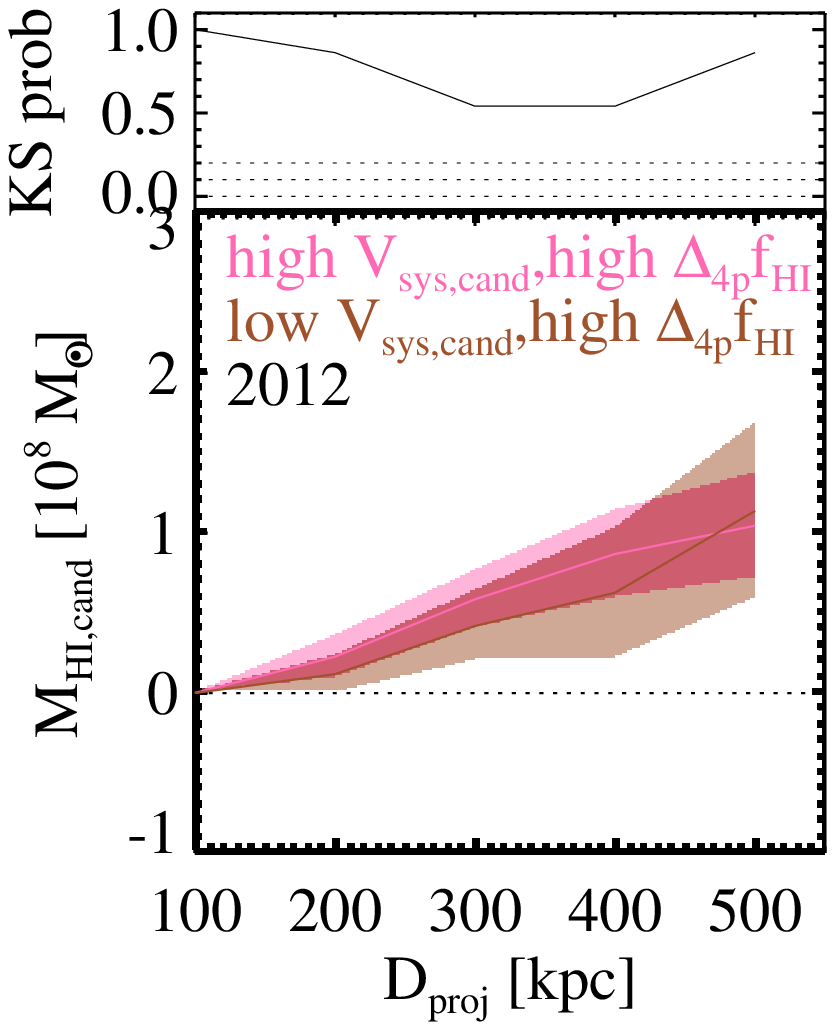}
\hspace{0.2cm}
\includegraphics[width=3.cm]{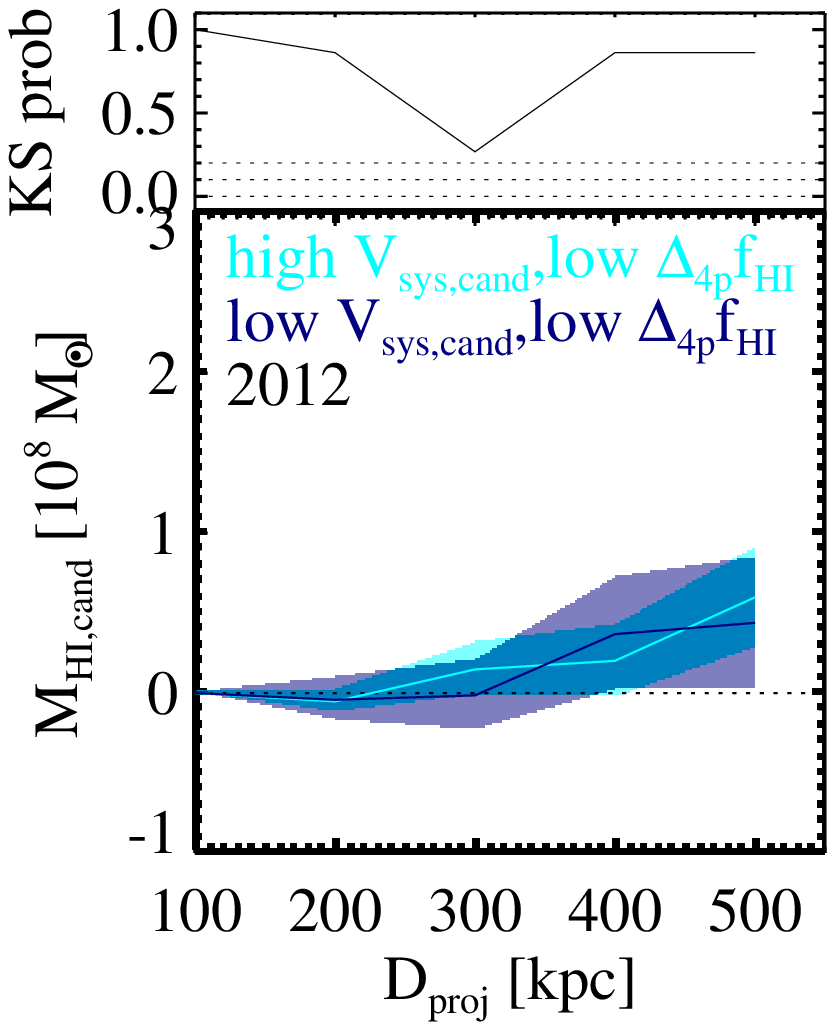}
\hspace{0.2cm}
\includegraphics[width=3.cm]{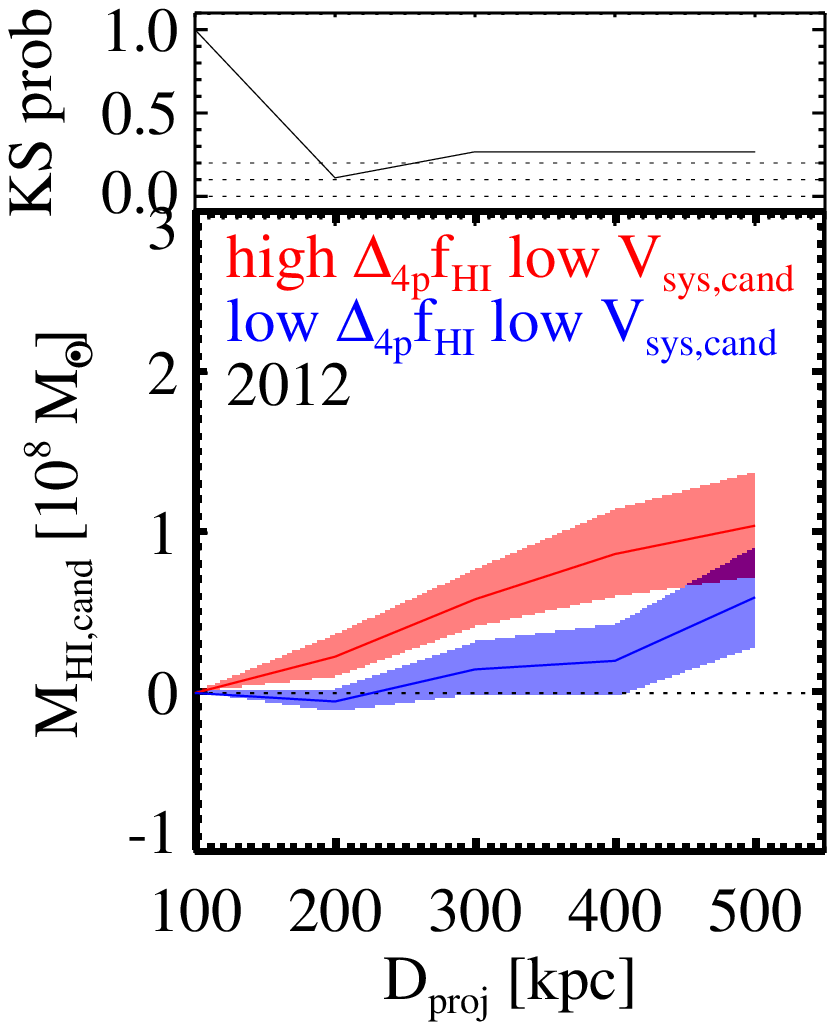}
\hspace{0.2cm}
\includegraphics[width=3.cm]{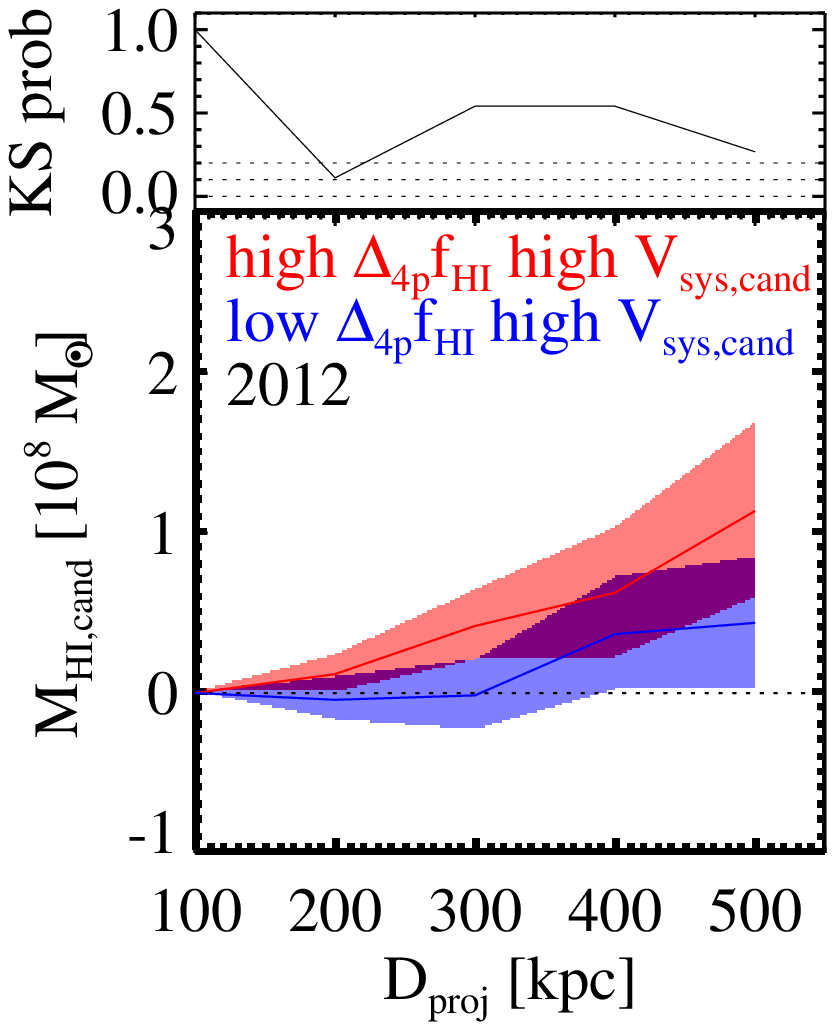}
\caption{ Similar to Figure~\ref{fig:HIprof}. The candidates are grouped according to their systematic velocity with respect to the primary galaxies, and the $\Delta_{4p}\fHI$ of the primary galaxies in the cubes.  The colors of the shades are for the sub-samples denoted in each panel. The top row is for the whole sample of 40 cubes and the bottom row is for the 28 cubes observed in the year 2012.}
\label{fig:app1}
\end{figure*}

In the main part of this paper, we report the detection of  \hi mass ($\mHIcand$) by adding the candidates outside the galaxies detected in \hi cubes (Sect. 3.3). We find $\mHIcand$ to be higher around galaxies with high \hi excess.
 We have performed tests to show that $\mHIcand$ is unlikely to be from CLEANing or continuum subtraction residual artefacts (Sect. 3.4). However, as shown in the top-left panel of Figure~\ref{fig:app1}, there is a small gradient in $\mHIcand$ between the blue and red shifted sides of the \hi cube  around the primary galaxies.   Consequently, part of the difference in $\mHIcand$ found between sub-samples defined by properties of primary galaxies may be related to this gradient.  Indeed, the gradient is  slightly stronger for the cubes with low $\Delta_{4p}\fHI$  primary galaxies than the cubes with high $\Delta_{4p}\fHI$ primary galaxies (the second and third panels in the top row of Figure~\ref{fig:app1}); and the difference of $\mHIcand$ around primary galaxies with low and high $\Delta_{4p}\fHI$  is less significant for the low V$_{sys}$ side of the cubes than for the high V$_{sys}$ side of the cubes.

\begin{figure}
\includegraphics[width=7cm]{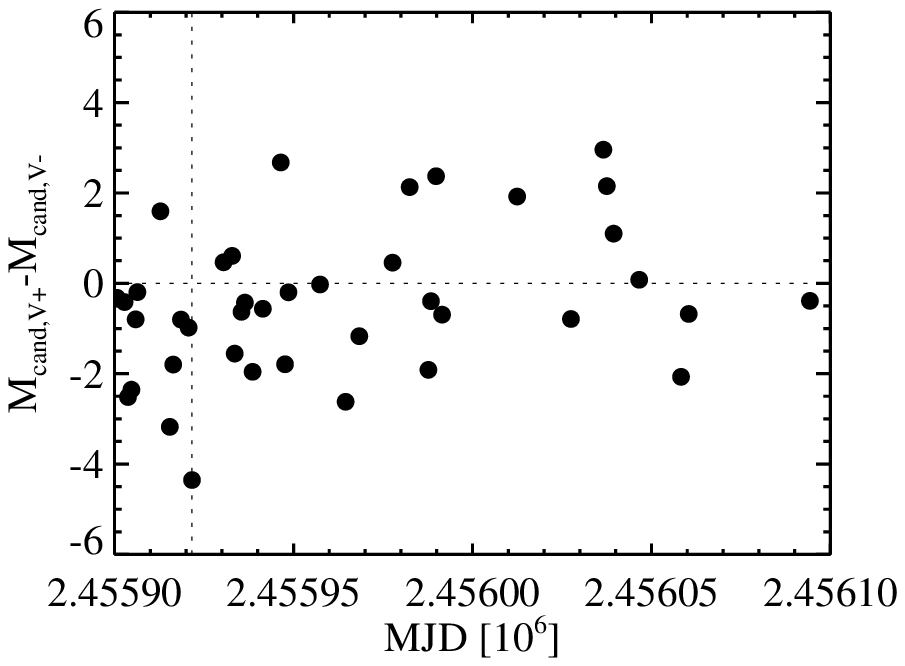}
\caption{  Relation between $\mHIcand$ gradients and Modified Julian Date (MJD).  The vertical dotted line marks the division of the year 2011 and 2012. The horizontal dotted line marks the position of 0. }
\label{fig:app2}
\end{figure}

From inspection of the position of  candidates in \hi cubes, we find that this gradient is not clearly associated with any failures of continuum subtraction, CLEANing  or RFI pattern removal. We do not find such gradients  in the test with stokes Q  cubes (Sect 3.4). However, we find this gradient is weakly correlated with the date of observation (Figure~\ref{fig:app2}) with a Pearson correlation coefficient of ~0.3, mostly driven by the data points for the year 2011, where the distribution of gradients is strongly biased toward negative values.
If we exclude the 12 data cubes observed in 2011, the gradients of $\mHIcand$ is much weaker for the whole sample, and for the primary galaxies with low and high $\Delta_{4p}\fHI$ as well (the first 3 columns in the bottom row of Figure~\ref{fig:app1}).  Meanwhile, the difference in $\mHIcand$ between the two sub-samples with low and high $\Delta_{4p}\fHI$ becomes more significant for the low V$_{sys}$ side of the cubes, and is similar for the high V$_{sys}$ side of the cubes (the fourth and fifth columns in the bottom row of Figure~\ref{fig:app1}).

Hence we argue that while the cause for the gradient of $\mHIcand$ is not found, the effect is not likely to significantly affect our main results,  Because it is unclear whether the gradient effect is an artefact or a statistic coincidence, we still use the full sample of 40 data cubes for analysis in the paper. 
Nevertheless, we warn the readers of this uncertainty in our experiment and intend to investigate it with larger samples in the future.  

\end{document}